\documentclass[table]{ccjnl}
\usepackage{lipsum,amsmath}
\usepackage{cuted}
\usepackage{bm}
\graphicspath{{figures/}}
\usepackage{adjustbox}
\usepackage{amsmath,amssymb,amsfonts}
\usepackage{algorithmic}
\usepackage{graphicx}

\usepackage{adjustbox}

\usepackage{makecell}
\usepackage{textcmds}

\usepackage{graphicx,booktabs}

\usepackage{multirow}

\usepackage{amsfonts}
\usepackage{lipsum,amsmath}
\usepackage{cuted}
\usepackage{bm}

\usepackage{textcmds}
\title{BEYOND 5G NETWORKS: INTEGRATION OF COMMUNICATION, COMPUTING, CACHING, AND CONTROL}

\author{Musbahu Mohammed Adam\inst{1}, Liqiang Zhao \inst{1,*}, Kezhi Wang\inst{2}, and Zhu Han\inst{3}} 

\receiveddate{TBD}
\reviseddate{TBD}
\Editor{TBD}

\address[1]{State Key Laboratory of Integrated Service Networks, Xidian University, Xi’an 710071, China }
\address[2]{Department of Computer and Information Sciences, Northumbria University, Newcastle upon Tyne, UK }
\address [3]{Department of Electrical and Computer Engineering, University of Houston, Houston, TX 77004 USA, and also with the Department of Computer Science and Engineering, Kyung Hee University, Seoul 02447 South Korea}

\begin{document}

\maketitle

\begin{abstract}
In recent years, the exponential proliferation of smart devices with their intelligent applications poses severe challenges on conventional cellular networks. Such challenges can be potentially overcome by integrating communication, computing, caching, and control (i4C) technologies. In this survey, we first give a snapshot of different aspects of the i4C, comprising background, motivation, leading technological enablers, potential applications, and use cases. Next, we describe different models of communication, computing, caching, and control (4C) to lay the foundation of the integration approach. We review current state-of-the-art research efforts related to the i4C, focusing on recent trends of both conventional and artificial intelligence (AI)-based integration approaches. We also highlight the need for intelligence in resources integration. Then, we discuss the integration of sensing and communication (ISAC) and classify the integration approaches into various classes. Finally, we propose open challenges and present future research directions for beyond 5G networks, such as 6G.
\keywords{4C; 6G; integration of communication, computing, caching, and control; i4C; multi-access edge computing (MEC)}
\end{abstract}

\section{Introduction}

Currently, there exist many research efforts in academia and industry that have been devoted to addressing the longstanding issues of communication, computing, caching, and control (4C) functionalities. However, a considerable number of these efforts focused on improving the performance of these underlying functionalities separately, which in turn leads to their unavoidable shortcomings. For instance, in the communication domain, the popular Shannon capacity limit is on the verge of being approached with the existing long-term evolution (LTE) techniques \cite{fan2016coping,wang2016power,zhang2017dynamics}. In the area of computing, the Moore's law is fast approaching its impending limit based on silicon chips technology. In the caching/storage domain, the rapid progress in magneto-optical and optical disks for storage may not accommodate the increasingly growing big data demands \cite{fan2016coping}. Likewise, the performance of control could be limited by multiple factors, including heterogeneous users’ demands, wireless fading channels, and insufficient computing power. In particular, wireless networked control systems share information among sensors, actuators, and plants using wireless networks, characterized by deep fade and susceptible to signal power loss. Such channel impairments render the control functionality/algorithm sub-optimal \cite{lima2020model}. Besides, control algorithm relies on extensive computations to run some networked control systems rapidly; thus, insufficient/weak processing units degrade the control performance. 

Due to these inherent limitations, further significant performance improvement in terms of communication, computing, or caching capabilities becomes more challenging for engineers and researchers in research and development sectors \cite{fan2016coping}. In other words, optimizing any one of the 4C functionalities/resources will hardly maximize the performance of a communication network \cite{zeng2020mobile}. Hence, relying on a single 4C functionality alone will no longer sustain the requirements of emerging intelligent applications and services. Nonetheless, the great advances in the individual domains of 4C triggered some promising steps toward proposing hybrid functionalities \cite{andreev2016exploring}, leading to the revolutionary changes that warrant the respective functionalities of 4C to encroach on one another's territory. Here is why one can hardly place a clear boundary among communication, storage (cache), computing \cite{fan2016coping}, and control domains nowadays.

\begin{figure*}[t]
	
	\centering
	\includegraphics[width=140mm]{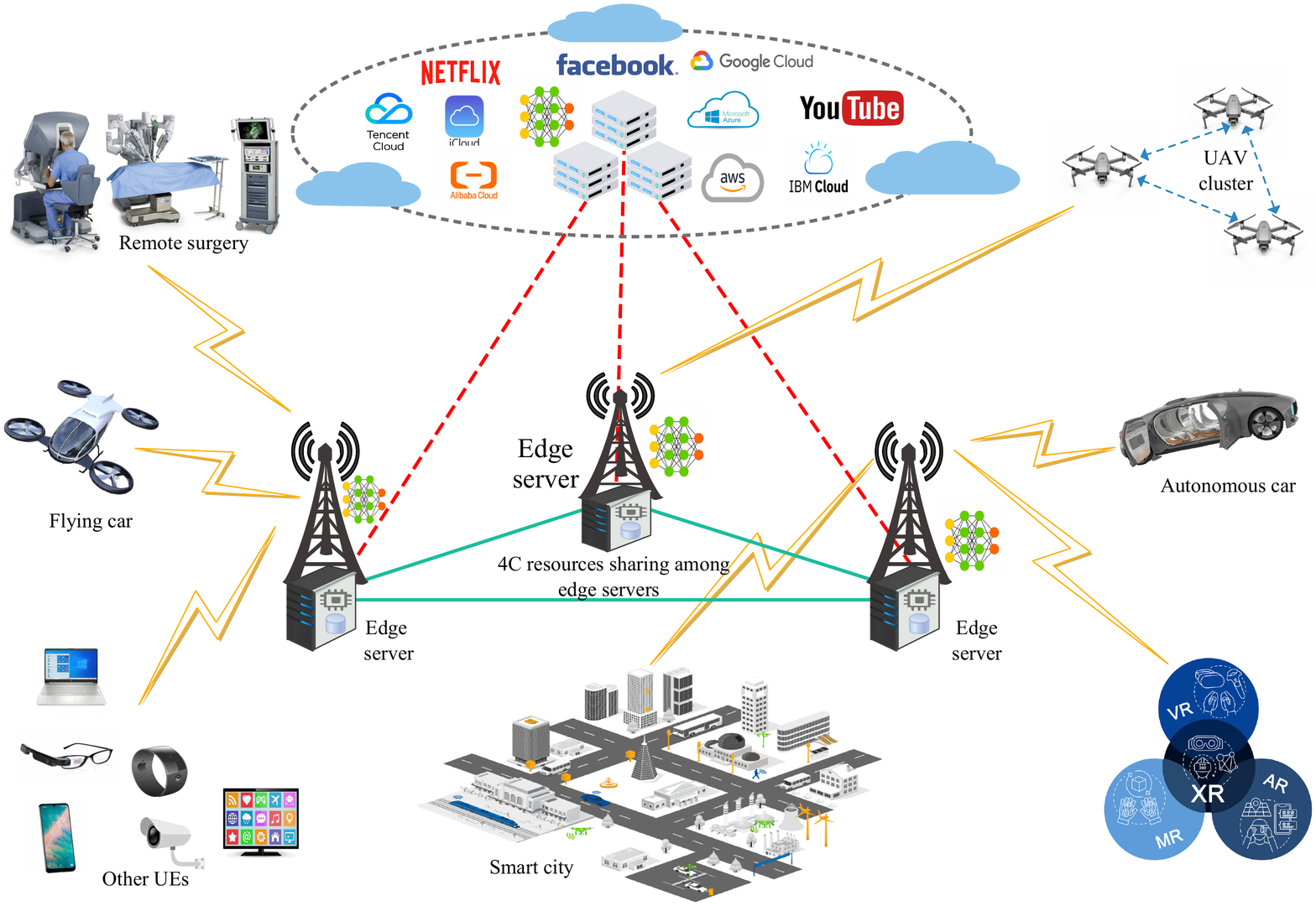}
	\caption{A typical scenario for the i4C in 5G, 6G, and beyond.}
	\label{fig2}
\end{figure*}

Moreover, the individual functionalities of 4C have great potential to complement and reinforce one another. For example, edge caching technique can minimize traffic redundancy, avoid duplicate transmission, and reduce bandwidth consumption in communications \cite{zeng2020mobile}. On top of that, the control algorithm is essential for controlling, coordinating, and optimizing the other integrants of 4C. The promising gains derived from the capabilities of 4C accelerate the progress toward integrating them in future networks. Achieving this striking breakthrough implies a paradigm shift toward the information transmission, processing, storage, and intelligent decision-making networks that support the new-technologies-new-services trends. Indeed, the integration of communication, computing, caching, and control (i.e., the i4C) will provide massive support for the fifth-generation (5G), sixth-generation (6G), and beyond networks, enabling key network elements, functionalities, and heterogeneous services.

\subsection{Integration for 5G, 6G, and Beyond}

Against prior network generations, the 5G network is emerging with a much more complex mission, i.e., supporting the dramatic evolution of information and communications technology (ICT) and the Internet. Hence, 5G systems support not only communication functionality but also the other three parts of the 4C functionalities. These functionalities play pivotal roles in enabling a variety of services in 5G, e.g., massive machine type communication (mMTC), enhanced mobile broadband (eMBB), and ultra-reliable and low latency communication (URLLC) \cite{mao2017survey,pham2020survey}. Today, with the advent of 5G networks, we witness a great boom in intelligent applications and use cases. Such applications keep emerging with unusual demands for communication, computing \cite{pham2020survey}, and caching resources. So far, the 5G's achievement in satisfying the stringent requirements of the applications is insufficient.

Furthermore, the evolution of 5G prompts the notion of beyond 5G networks, such as 6G and beyond, for superior performance. In 6G networks, novel disruptive wireless technologies and futuristic network architectures will be put into perspective. 6G is further envisaged to ultimately attain future generation connectivity, driven and motivated by the transformation from \qq{\textit{connected everything}} to \qq{\textit{connected intelligence,}} hence facilitating \qq{\textit{human-thing intelligence}} interconnectivity \cite{bariah2020prospective}.

Therefore, compared with 5G network, 6G is expected to surface with larger dimensions, higher complexity, dynamicity, and heterogeneity features. These issues require a novel, agile, flexible, adaptive, and intelligent architecture. Featured with intelligent recognition, high learning, predicting, and powerful reasoning and decision-making abilities, \textit{artificial intelligence (AI)} can allow the 6G network architecture to learn, intelligently decide, and adjust itself in order to enable diversified services without requiring human support \cite{yang2020artificial}. To this end, Letaief et al. \cite{letaief2019roadmap} applied AI techniques to realize network intelligentization, closed-loop optimization, and intelligent wireless communications for 6G networks.

Moreover, AI will serve as a powerful assistant for the communication, computing, and caching functionalities in edge computing. This promising novel paradigm is termed \textit{intelligent edge}. With the AI techniques, intelligent edge provides optimal edge computing solutions, such as resource allocation optimization \cite{xu2020edge}.

The idea of softwarization alone will no longer be sufficient for the 6G mobile networks due to the growing complexity and heterogeneity in mobile wireless communication networks. To be more specific, network elements should support different capabilities, comprising AI control, communication, computing, content caching, and even wireless power transfer, for supporting mobile AI-based applications \cite{letaief2019roadmap}. In other words, converging intelligent control, sensing, communication, computing, and caching functions will be a core driver behind the emerging 6G networks. The recent influx of the AI frontiers (e.g., deep learning (DL), federated learning (FL), machine leraning (ML), and deep reinforcement learning (DRL) techniques) at the network edges implies the urgent need for the intelligent decision-making techniques that can efficiently interact the conflicting functionalities in future networks. In fact, 6G networks will rely on the AI frontiers to realize intelligent optimization of 4C. Beyond that, integrating communication, computing, and caching with AI-based/intelligent control could be one of the biggest revolutionary trends that come to stay in 6G and beyond mobile networks for disruptive applications/innovations.

In short, 6G networks will unleash the maximum potential of communication with computing and control at the proximity of myriad mission-critical applications \cite{letaief2019roadmap}, such as extended reality (XR), Tactile Internet, autonomous vehicles, Internet of Vehicles (IoV), Internet of Everything (IoE), Internet of Intelligent Things (IoIT),  flying vehicles, and space-air-ground integrated networks. The i4C will indeed allow the 6G mobile networks  to make appropriate decisions on the user equipment (UE) applications' tasks. The control will ensure that only tasks to be completely and optimally executed, given the available resources at the moment, are granted resources. Hence, the convergence of 4C will bring unprecedented solutions to multitudinous intelligent applications that emerge with stringent requirements for massive connectivity, ultra-high reliability, ultra-low latency, high mobility, energy-saving, and so on. Above all, the i4C will drive wireless networks to reach newer and greater heights of performance. Fig. 1 portrays a typical scenario where the integrated 4C resources embedded in distributed collaborative network edge servers are shared for different quality of service (QoS) requirements. Therefore, moving toward integrating 4C (i.e., i4C) should be the focal point for the evolution of 6G and beyond mobile networks.

\subsection{Motivation and Contributions}

%%%%%%%%%%%%%%%%%%%%%%%%%%%%%%%%%%%%%%%%%%%%%%%%%%%%%%%%%%%%%%%%%%%%%%%%%%%%%%%%%%%%
\begin{table*}[t]
	\caption{Comparison of existing surveys on the integration of resources}
	\label{table2}
	\begin{center} 
		
		{\tiny\renewcommand{\arraystretch}{1}
			\resizebox{!}{.17\paperheight}{%
				
				\begin{tabular}{ | p{0.7cm} | p{1.5cm} | p{3.7cm} |p{2.7cm} |}
					\hline

					\hline
					
					\hline \hline

					\textbf {Related Surveys}	 & \textbf{Themes } &  \textbf{Key Contributions} & \textbf{Limitations}\\

					\hline 
					
					\hline
					
					\hline  
					%%%%%%%%%%%%%%%%%%%%%%%%%%%%%%%%%%%%%%%%%%%%%%%%%%%%%%%%%%%%%%%%%%%%%%%%%%%%%%%%%%%%%%%%%%%%%
					
					%%%%%%%%%%%%%%%%%%%%%%%%%%%%%%%%%%%%	
					
					\cite{mao2017survey}  &  Communication and Computation Resource Allocation  & $\bullet$ Surveyed communication and computation models in MEC. \newline $\bullet$Reviewed communications and computation resources allocation.& $\bullet$Omitted caching and control and ignored their models. \newline $\bullet$ Omitted recent efforts on AI and edge intelligence. \\ 
					
					\hline

					%%%%%%%%%%%%%%%%%%%%%%%%%%%%%%%%%%%%

					\cite{pham2020survey}  &  Integrating MEC into 5G Technologies & $\bullet$	Focused on fusing MEC and 5G technologies. \newline $\bullet$ Applications of ML in MEC, comprising 4C optimization, big data, etc. & $\bullet$ Discussed 4C optimization as an application of ML in MEC.
					\newline $\bullet$ Omitted 4C models.
					\\ 
					\hline 
					
					%%%%%%%%%%%%%%%%%%%%%%%%%%%%%%%%%%%%	

					\cite{wang2017integration,wang2017survey,bouras2020convergence}  &  Converging Communications, Computing, and Caching in Mobile Edge Networks  & $\bullet$ Reviewed issues of converging/integrating communication, computing, and caching.\newline $\bullet$ Focused on definition, architecture/frameworks, enablers, metrics, IoT, and challenges.
					& $\bullet$ Did not cover recent efforts on edge intelligence and AI solutions for i4C.
					\newline $\bullet$ Omitted control.
					\newline $\bullet$ Ignored 4C models.
					
					\\ 
					\hline

					%%%%%%%%%%%%%%%%%%%%%%%%%%%%%%%%%%%%	

					\cite{kim2012cyber}  &  Integrating Communication, Computing, and Control  & $\bullet$ Surveyed the integration of communication, computing, and control.
					\newline $\bullet$ Focused on real-time computing, networked control, real-time networking, wireless sensor networks, and autonomous vehicles.
					& $\bullet$ Ignored efforts on network edges, edge intelligence, and the AI roles in i4C.
					\newline $\bullet$ Did not cover caching.
					\newline $\bullet$ Omitted 4C models.
					
					\\ 
					
					\hline
					%%%%%%%%%%%%%%%%%%%%%%%%%%%%%%%%%%%%
					
					This \newline Survey  &  i4C  & $\bullet$ Reviewed the i4C with emphasis on AI-based and conventional integration approaches. 
					\newline $\bullet$ Surveyed the integration aspects, including motivations, key enablers, applications and use cases.
					\newline $\bullet$ Discussed 4C models, challenges, and future direction.
					&
					
					\\

					%%%%%%%%%%%%%%%%%%%%%%%%%%%%%%%%%%%%
					
					\hline 
					
					\hline
					
					\hline  \hline
					
				\end{tabular}
				
		}}
	\end{center}
	
\end{table*}

%%%%%%%%%%%%%%%%%%%%%%%%%%%%%%%%%%%%%%%%%%%%%%%%%%%%%%%%%%%%%%%%%%%%%%%

In recent years, we witness many promising gains brought by the evolution of information technology (IT) in wireless network environments. The appreciable advances in IT and communications have been revolutionizing conventional networks in terms of structures and operations, driving new capabilities by leveraging synergies among different functionalities of 4C. Considering significant benefits brought by the individual functionalities of 4C, integrating them into a single system/network becomes necessary for users' satisfaction and networks' performance requirements. Converging 4C may lead to realizing the optimal solutions that will fulfill diverse QoS requirements of futuristic intelligent applications and use cases in the coming decades. Suffice to say, the i4C becomes a natural trend that has pivotal roles to play in 5G, 6G, and beyond networks.

Today, different aspects of 4C receive attention from academia and industry. In fact, several research efforts investigated the trends and issues of converging communication, computing, and caching and that of communication, computing, and control. However, a few survey articles, including \cite{mao2017survey}, \cite{pham2020survey}, and \cite{wang2017integration,wang2017survey,bouras2020convergence,kim2012cyber}, focused on these concepts. Table 1 summarizes the contributions of these surveys. Specifically, Mao et al. \cite{mao2017survey} reviewed efforts on mobile edge computing (MEC), focusing on joint computation and communications resource allocation in MEC and their models. Pham et al. \cite{pham2020survey} focused on integrating MEC with potential 5G technologies and discussed applications of machine learning in MEC, comprising 4C optimization, crowdsensing, big data, and privacy and security. The survey in \cite{wang2017integration} reviewed recent works on mobile network edges, exploring issues of communication, computation, and caching techniques, and discussed challenges and applications of edge networks. Wang et al. \cite{wang2017survey} reviewed some efforts intended to integrate communication, computing, and caching. The central theme of the survey focused on key aspects of the integration, i.e., motivations, enabling technologies, performance metrics, frameworks, and challenges. In \cite{bouras2020convergence}, the authors reviewed the trends of technology in communication, caching, and computing resources, analyzing the interactions among the resources while collecting, storing, indexing, and processing IoT data. They also described the convergence in devices, sensors, and gateways. \cite{kim2012cyber} focused on integrating communication, computing, and control and its potential applications.

\begin{figure*}[t]
	\centering
	\includegraphics[width=18cm, height=10cm]{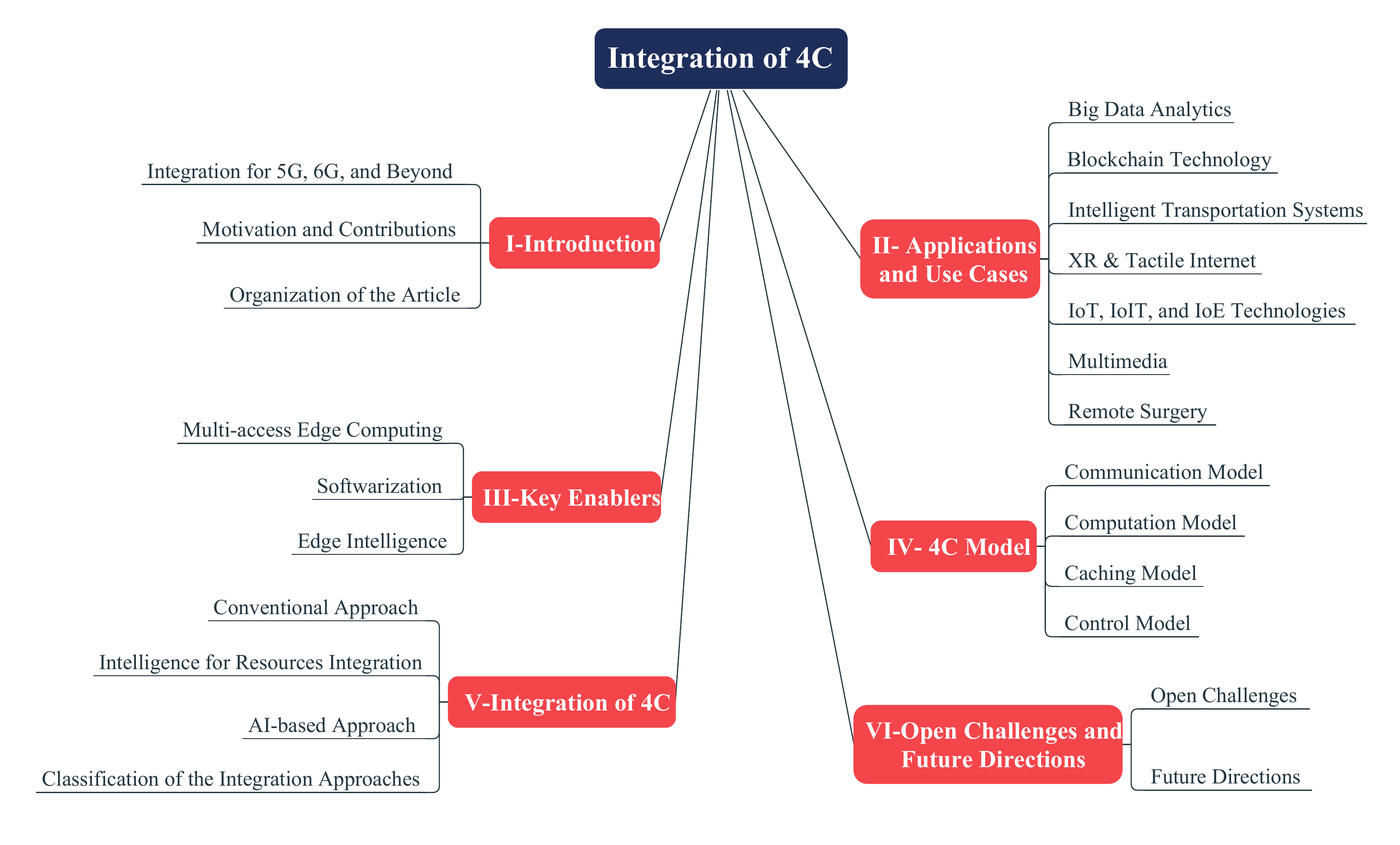}
	\caption{The roadmap of the survey.}
	\label{fig3}
\end{figure*}

These surveys contributed greatly to optimizing the networks' performance and improving the users' experience. Nevertheless, neither integrating communication, computing, and caching nor integrating communication, computing, and control is sufficient. Of course, the roles of control in resources integration cannot be overstated. For example, in radio access network (RAN) slicing and distributed resource allocation, a control scheme has to be put in place to guide the allocation and deallocation of the competitive network resources. Besides, to satisfy diverse QoS needs, 5G New Radio (NR) is designed to be flexible; the 6G network is expected to be much more flexible and complex. The growing flexibility in these networks implies more control parameters necessitating essential changes in wireless network operations \cite{challita2020machine}. Likewise, caching saves network bandwidth and avoids the transmission of duplicate content. Therefore, the need for i4C arises to shape up the visions of future generation networks.

To the best of our knowledge, there is no existing survey that devoted itself to converging 4C. Hence, this paper aims to present a firsthand tutorial on the convergence of 4C against the aforementioned surveys. The survey differs from the previously mentioned ones with the following contributions. To begin with, the paper discusses different aspects of the integration, exploring its background, motivations, leading enabling technologies, and potential benefits and use cases. Another point worth noticing is that the existing surveys omitted the 4C models, which serve as the backbone of the design and implementation of an integrated 4C network/system. Considering the roles of AI in 6G networks and the increasingly growing complexity of wireless networks, the paper pays much attention to the recent trends of i4C based on the AI techniques. Thus, the paper comprehensively reviews the i4C, focusing on conventional and AI-based approaches. It also classifies various approaches of the integration and discusses the integration of sensing and communication (ISAC). Then, it considers several open challenges and provides future directions. The roadmap of this survey is shown in Fig. 2.

\subsection{Organization of the Article}

The rest of this paper is organized as follows: Section II brings a snapshot of potential applications and use cases. Section III presents the main enablers for i4C in future mobile networks. Section IV discusses various models of 4C, which lays the foundation for their integration. Section V reviews many cutting-edge research efforts on the i4C with a focus on both AI-based and conventional optimization/integration approaches. It also  discusses the convengence of communication and sensing and classifies different approaches of integrating resources. Section VI focuses on open challenges and explores future research directions. Finally, Section VII concludes the survey.

\section{POTENTIAL APPLICATIONS AND USE CASES}

Numerous drivers motivate the convergence of 4C functionalities. One of the key driving forces behind the i4C is the explosion of wide-ranging intelligent applications and use cases, including big data, autonomous cars, telesurgery, Internet of Things (IoT),  XR \& Metaverse, Tactile Internet, multimedia, and energy/power systems. Generally, such applications come up with different service requirements, including ultra-low latency, higher throughput, ultra-high reliability, intensive computational capabilities, and vast caching resources. This section highlights the roles of i4C in accommodating the needs of these applications.

\subsection{Big Data Analytics }

Data analytics undergoes revolution in numerous scientific domains due to the exponential growth of data. This complex and enormous volume of data calls for parallelization at an unprecedented scale because processing it may exceed the capabilities of a single or even a couple of machines \cite{zhang2016collective}. Today, mobile UEs and other IoT devices offload tasks and corresponding data via wireless channels with varying rates, and such data, owing to its diversity, scale, and timeliness,  may require real-time analytics and live stream computing and caching. These raise the need for fast, parallel, and distributed processing \cite{ndikumana2019joint}. To this end, 4C is pushed closer to the devices generating the data, avoiding the bottlenecks resulting from the need of moving data from the storage to the central processing unit (CPU) and its main memory and back. In this way, real-time response, lower latency, and energy saving can be guaranteed \cite{torabzadehkashi2019catalina}. 

To support big data analytics, characterized by increasingly growing volume, variety, and velocity (3Vs) of data, the big data infrastructure has to couple a scalable data storage with an ultrafast data processing system \cite{zhang2017quantcloud}. Thus, the analytics system should be configurable, flexible, and scalable both vertically and horizontally \cite{maarala2015low}. Of course, a highly distributed collaborative system of 4C deployed at the network edges will serve significant roles for big data requirements in future networks. Furthermore, the convergence of 4C capabilities at the network edges can allow 6G networks to handle huge volumes of data with high-speed data rate connectivity per UE \cite{chowdhury20206g}. Fig. 3 shows a collaborative system of MEC servers hosting 4C resources for processing big data. In a word, the i4C is a potentially promising approach to cope with the big data challenges in 5G and 6G networks.

\subsection{Blockchain }

\begin{figure}[t!]
	\centering
	\includegraphics[width=90mm]{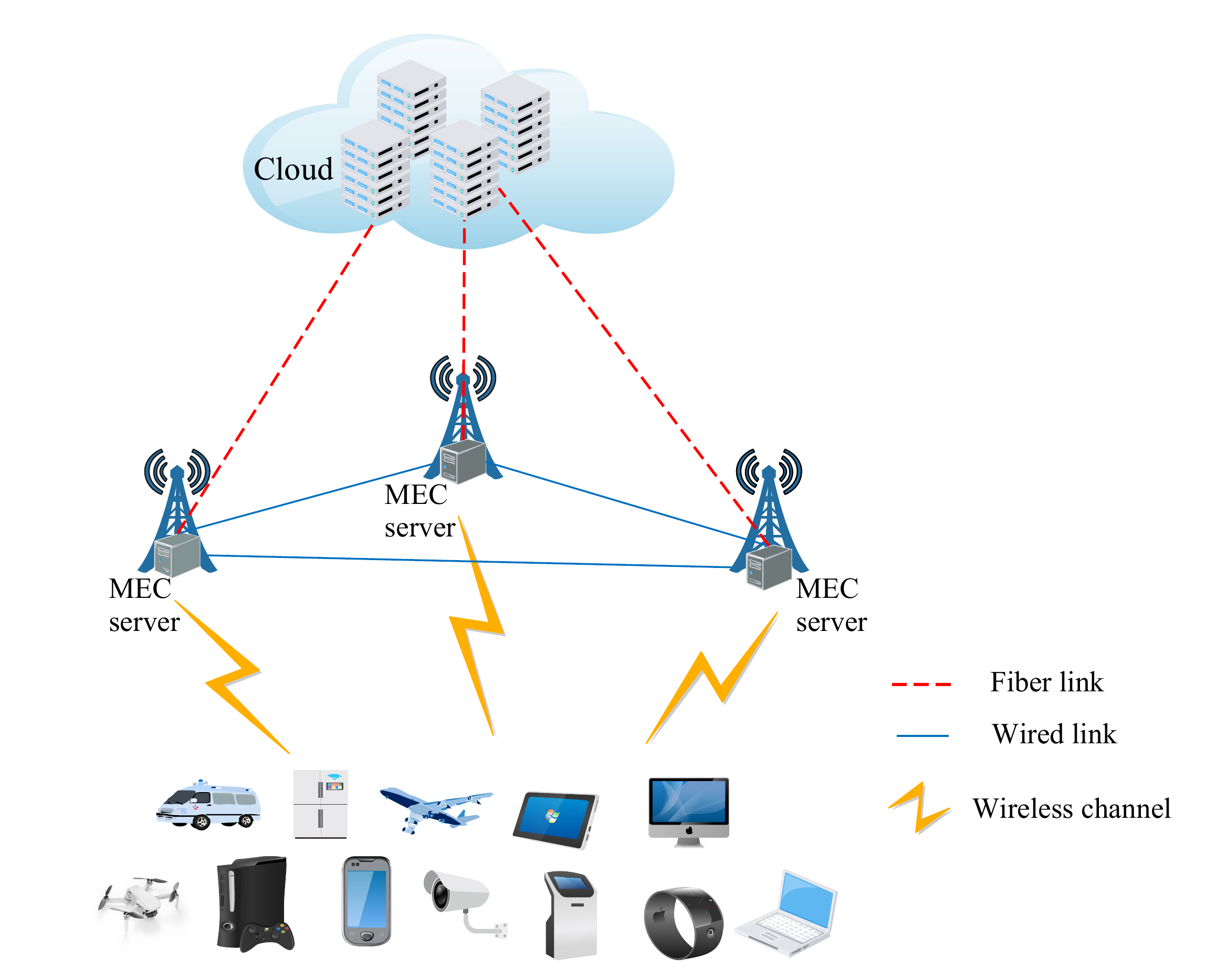}
	\caption{Collaborative MEC servers for big data processing.}
	\label{fig4}
\end{figure}

Besides its bright potential for implementing key services in 5G, 6G, and beyond networks, blockchain technology deserves recognition for its remarkable achievement in cryptocurrencies, such as Bitcoin, Cardano, Ethereum, and other trending metaverse applications. However, due to their distributed nature, such cryptocurrencies call for high bandwidth, sensing, computing, and caching capabilities for pledging the ledger integrity. The cryptocurrency protocol handles huge volumes of data transmitted/broadcasted across the participating/playing nodes \cite{jahid2021convergence}.

Despite its enormous potential, blockchain faces some critical issues, including scalability and latency, that limit the performance of its applications \cite{jahid2021convergence,jiang2021road}. To realize a scalable blockchain system with real-time applications, the 4C functionalities can be tightly converged in distributed systems to handle the demands for ubiquitous connectivity, caching, and computing capabilities.

\subsection{Intelligent Transportation Systems }

Recently, autonomous vehicles begin to surface with the advent of 5G networks. The key performance requirements of such vehicles include ultra-high reliability and lower transmission delay in terms of millisecond scale. However, the dynamic on-vehicle information processing rates and the randomness of wireless communication channels limit the performance of autonomous vehicles. Due to these inherent limitations, the vehicle-to-vehicle (V2V) communication links will unavoidably experience time-varying delays. Using delayed information in designing the control system of autonomous vehicles could jeopardize the stability of the platoon system \cite{zeng2019joint}. This issue requires a robust integrated system of 4C  to sustain the stability of the platoon.

In future transportation systems, intelligent infrastructures and intelligent vehicles will converge. By operating the intelligent traffic infrastructures, the whole traffic systems' throughput can be managed efficiently. On the other hand, by equipping intelligent vehicles with seamlessly integrated systems of embedded computing and in-vehicle networks, wireless data exchange can be enabled between vehicle-to-infrastructure and vehicle-to-vehicle. These two capabilities can allow the vehicles to guide drivers or even drive independently (autonomously) by observing and evaluating traffic conditions, planning ahead of their behavior, and actualizing (implementing) the plan using the drive-by-wire functionalities, which include steering, speed and stability controls, braking, and so on \cite{kim2012cyber}. Hence, the i4C becomes necessary for sustainability, ultra-low latency, efficiency, ultra-high reliability, stability, and safety of autonomous vehicles. In 6G and beyond networks, the convergence of 4C will undoubtedly be fully leveraged to realize the visions of fully autonomous vehicles and the IoVs.

\subsection{XR \& Tactile Internet }

XR is an umbrella term referring to the set of immersive technologies, comprising AR, VR, and mixed reality (MR). To transmit higher resolution/frame rate videos, mobile AR/VR applications impose high demands for greater bandwidth with ultra-low latency, ultra-high reliability, and high computing power. Likewise, the Tactile Internet applications, which may extend to healthcare, entertainment, robotics, and autonomous vehicles, require higher communication bandwidth with ultra-low latency. Thanks to the nature of haptic signals and human perception, the required latency for Tactile Internet is 1 ms. Unfortunately, satisfying the stringent requirements of these applications is beyond the capabilities of the existing LTE networks \cite{sukhmani2018edge}. The 5G and 6G wireless networks will support the tight convergence of 4C at the vicinity of the users' applications. This implies promising solutions for latency-sensitive applications. Therefore, with an integrated system of 4C at the network edges, the requirements of both Tactile Internet and XR applications will be fulfilled.

\subsection{IoT \& IoE Technologies}

One of the intrinsic features of IoT is its potentiality to provide users with in-built intelligence by enabling its devices (aka IoT devices) to connect. Most of the resource-hungry IoT devices interact with one another, reforming the way individuals perceive their environment and get information. Such smart devices can sense and access information from their surrounding environment and accordingly form what is termed sensory swarm. The accessed information may be conveyed to different applications for processing and analysis \cite{wang2018hierarchical}. The IoT devices share and interpret information based on standardized formats, and by leveraging the essential functionalities of 4C, IoT transforms its devices from traditional to smart. Hence, the i4C will impact the advancement of flexible and efficient IoT in smart cities, providing end-users with diverse smart services thereby enriching energy, entertainment, environment, healthcare, and transportation \cite{bouras2019synergy}. Furthermore, the impact of i4C will potentially reach beyond the concept of IoT. Of course, the emerging IoE paradigm, where everything is connected to everything, requires the convergence of 4C; likewise, the concept of IoIT.

\subsection{Multimedia}

The current prevalence of UEs results in the rapid growth of the internet traffic \cite{gao2017vcache}, which is dominated by video streaming. Today, video streaming accounts for more than 70\% of North American downstream traffic at peak time. However, with unstable wireless network conditions, insufficient bandwidth, and billions of viewing devices, the user experiences are inherently deteriorated, sparking a tussle between the increasing video traffic demand and the quality of viewing experiences. In such conditions, adaptive bit rate (ABR) streaming can be used to enhance viewing experiences. Nevertheless, ABR requires tremendous caching and computing resources for pre-transcoding of each video and caching all video chunks \cite{Ahmed2016ASO}. Thus, future mobile networks have to integrate 4C functionalities in the vicinity of the UEs in order to enable efficient multimedia service delivery (in terms of seamless connectivity, lower latency, and high reliability).

\subsection{Remote Surgery} 

In the coming decades, the 6G networks and beyond will be providing improved healthcare services for human beings. Remote surgery/Telesurgery enables a doctor to perform a surgical operation on a patient without being in the same physical location. This requires ultra-high-speed data rates, ultra-low latency, ultra-high reliability, flexibility, high computational power, and the rest, which can hardly be guaranteed by the 5G capabilities. In 6G and beyond, the 4C functionalities will integrate to steer the remote surgery to greater heights. Moreover, many future applications and use cases, among others holographic teleportation, intelligent production, and intelligent life, will undoubtedly require the i4C.

\section{KEY ENABLERS FOR INTEGRATING 4C}

This section focuses on the key enabling technologies for the i4C in 5G, 6G, and beyond networks. Specifically, Section III-1 gives a snapshot of multi-access edge computing, Section III-2 discusses softwarization, and Section III-3 focuses on edge intelligence. These fundamental enabling technologies have, indeed, proven their capabilities in providing diversified network solutions.

\subsection{Multi-access Edge Computing (MEC)}

MEC emerged to bring the IT and cloud computing capabilities into the RAN domain \cite{mao2017survey}. Due to the emergence of MEC, 4C is moved to mobile network edges nowadays \cite{mao2017survey}, \cite{ndikumana2019joint,ndikumana2017collaborative}. With the 4C functionalities at the network edges, data and computational tasks can be wirelessly offloaded, analyzed, computed, and stored near the UEs. In this way, both energy consumption and end-to-end latency will be significantly reduced. However, integrating MEC with a wireless network environment poses some challenges about the control and coordination of joint communication, computing, and caching \cite{ndikumana2019joint}. Hence, there is a need for optimal decision making on both networks and UEs’ computational tasks/data. 
In the design of an efficient MEC framework, it is essential to flawlessly couple computation offloading control and communication resource management in order to adapt the random variations of wireless channels in terms of frequency, space, and time \cite{mao2017survey}.

On the other hand, the variations in both wireless channels and available computing resources necessitate the need to intelligently control the input data size in UEs for local computing and efficient computation offloading. In doing so, the overall energy consumption for local CPU and transmission will be reduced under a task-deadline constraint \cite{tao2019stochastic}. Therefore, joint control of local computing and computation offloading has to be considered for swift execution and caching of the computational tasks. Fig. 4 shows a typical MEC system.

\begin{figure}[t]
	\centering
	\includegraphics[width=9cm]{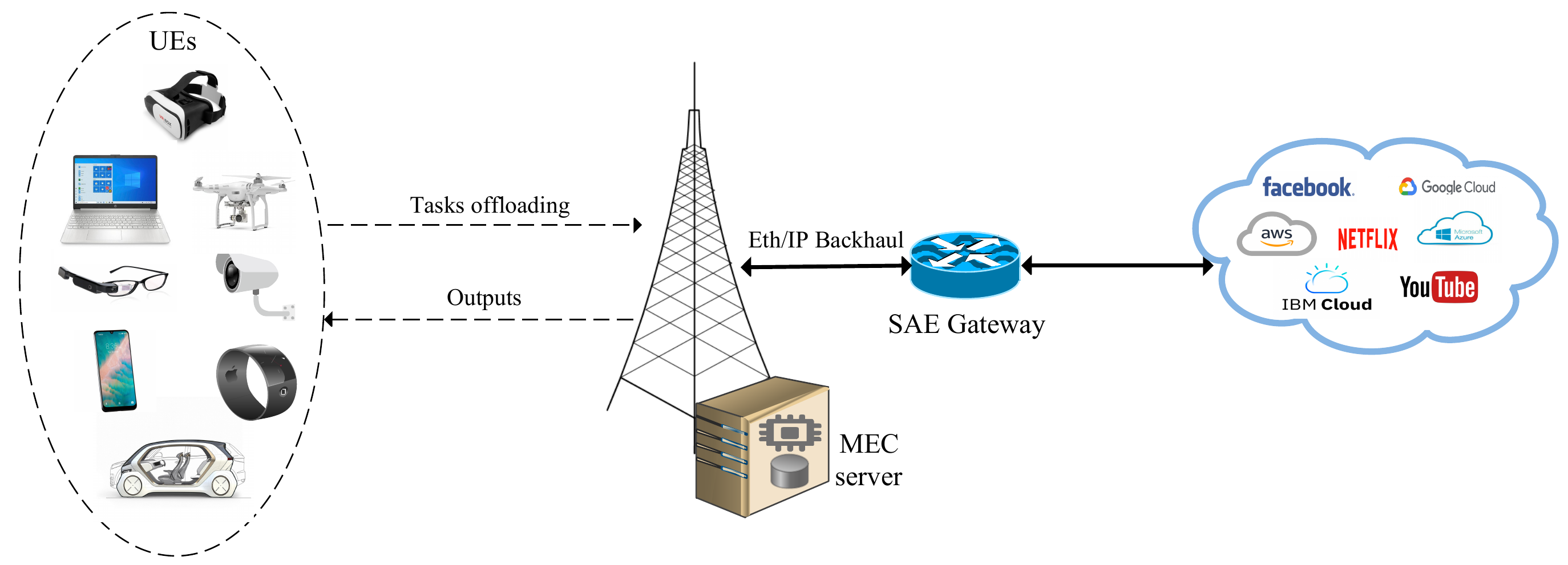}
	\caption{A typical MEC system.}
	\label{fig}
\end{figure}

Besides, both computing and cache resources rely on the available communication resources. Specifically, the communication resource is required to offload data and tasks for processing, analyzing, and caching at the network edges \cite{ndikumana2019joint}. In a 5G mobile edge network, a smart base station (SBS), widely considered a primary 5G infrastructure, will enable the integrated 4C services. This requires leveraging the ultimate synergy of the integrated caching, communication, and computing operations by comprehensively considering all essential control factors. In other words, an integrated control scheme is needed to harness the existing synergies between the communication, computing, and caching capabilities for realizing the optimal performance of the 5G network. Thus, in the 5G edge network, a control functionality has to interact the rational agents with conflicting objectives in the SBSs \cite{kim20185g}. Accordingly, \cite{chen2018joint} emerged with a framework that programmably controls and integrates in-network caching, networking, and computing for essential network operations. This further entails the necessity of converging 4C in 5G, 6G, and beyond mobile networks.

\subsection{Softwarization}

Network functions virtualization (NFV), software-defined networking (SDN), and information-centric networking (ICN) make the leading candidates for softwarization \cite{cabrera2018softwarization}, playing significant roles in the i4C. The NFV adopts virtualization techniques to flexibly program the network service functionalities as software instances, i.e., virtual network functions (VNFs), at the network edge servers. The MEC and NFV converge at the network edge to enable the provisioning of computation-oriented services. Thus, various compute-intensive applications will be greatly supported, thereby reducing both operating and capital expenses. Conversely, SDN decouples the control and data planes to improve the network-layer data traffic forwarding and optimizes the network-level resource orchestration; it utilizes a centralized controller in the control plane to receive the network information. The controller, having a global view of the network, makes the network-level decisions for resource allocation, access control policies for UEs, and traffic routing path configuration among the network components, which include network edge servers, access points/BSs, and network switches in the RAN and core network, for enhancing QoS and improving overall resource utilization \cite{zhuang2019sdn}. Other than splitting the control and data planes, SDN follows the abstraction principles, comprising data traffic forwarding, routing, and configuration as a computing problem. Employing these abstractions results in enabling the network slicing functionalities, and slicing the network leads to enabling the network resource allocation \cite{cabrera2018softwarization}.

Hence, the network resource allocation can be effectively managed and optimized by SDN. Adopting SDN to serve as a control module for the integration of resources becomes natural due to its efficiency and effectiveness in managing wireless networks. Nevertheless, the concept of control and resource allocation mechanism of the unified 4C solution is wider than the concept SDN alone; likewise, its diversity goes beyond the reach of any single technology \cite{wang2017integration}. This implies the need for hybridizing enabling technologies to realize the i4C solutions in future networks.

5G network requires an integrated approach composed of MEC, cloud, and core network. In this paradigm shift, NFV and SDN have challenging roles to play in transforming the way of managing wireless networks \cite{cabrera2018softwarization}. Coupling an NFV and SDN framework with MEC brings centralized network control over communication, computing, and caching resources, thus improving multi-resource orchestration efficiency \cite{zhuang2019sdn}. In short, integrating NFV and SDN promises flexible network infrastructure, resource management as well as new applications deployment \cite{zhang2018software}.   

On the other hand, ICN enables content retrieving for UEs based on identifiers (content identification), not on the basis of physical locations. In 5G, 6G, and beyond, ICN will serve similar roles as NFV and SDN \cite{cabrera2018softwarization}. Note that these promising networking paradigms do not compete, they complement one another instead; they handle various networking issues while benefitting one another. For example, ICN and SDN can be combined to form an SD-ICN framework, hence realizing holistically optimal resource allocation through the logically centralized controller. Moreover, integrating SDN and ICN brings several gains, solving the host-centric networking (TCP/IP) problems and tackling the caching and control issues; see \cite{zhang2018software}.

Today, ICN, NFV, and SDN converge to provide promising 4C solutions. ICN offers a new approach for provisioning services in SDN and NFV; it can also virtualize the network edge functions with some degree of data plane programmability \cite{zhang2018software}. Due to its ability to allow better migration to cutting-edge technologies through isolation of network parts, SDN-based virtualization excels as a decent approach for converging heterogeneous networks (Het-Nets) with MEC and ICN \cite{zhou2017resource}, \cite{zhou2018communications}. It brings several benefits: i) it allows flexible management and maintenance of Het-Net framework; ii) with the abstraction and standardization of the data and control planes, networks and applications can be evolved and updated without redesigning the network infrastructure; iii) the control plane can be pushed to the edge/cloud servers rather than a dedicated platform; and iv) both operation and capital expenditures are reduced by utilizing advance software and conventional hardware \cite{zhou2018communications}. In \cite{zhou2018communications}, an integrated resources mechanism was built upon the concept of SDN and wireless network virtualization, where MEC and ICN reinforce each other for promoting network efficiency and guaranteeing diverse service requirements.

In 5G and 6G networks, the promising capabilities of ICN, NFV, and SDN will be leveraged to tightly integrate 4C to meet the QoS needs of diverse applications. However, softwarization alone is insufficient for 6G due to the increasing complexity and heterogeneity of wireless networks \cite{letaief2019roadmap}. This opens a new avenue for integrating communication, computing, and caching with intelligent control in 6G and beyond. Here is where the concept of \textit{edge intelligence} and \textit{intelligent edge} begins.

\subsection{Edge Intelligence}

One of the essential network entities missed in the 5G mobile networks is edge intelligence, powered by the AI frontiers. Edge intelligence will, in all likelihood, serve the role of a key component in 6G networks to enable new functions, services, and superior performance \cite{peltonen20206g}. Today, edge intelligence is increasingly becoming a center of attention, attracting several research endeavors, due to its promising future and great benefits in 6G and beyond networks.

There are synergistic benefits between the AI techniques and edge computing. For example, edge computing unleashes its scalability and potentials with AI, and the AI technique allows innovations and algorithms for edge computing. Besides, AI extends its applicability to edge computing, and edge computing offers scenarios and platforms for AI. Thus, the AI techniques and edge computing will support and reinforce each other. The prospect of combining edge computing and AI has triggered a solid interest in both academia and industry. The integration of AI and edge computing , which is considered natural and unavoidable, results in the birth of edge intelligence. Actually, edge intelligence goes beyond a mere fusion of the AI techniques and edge computing. The concept of edge intelligence is wide enough and greatly sophisticated, covering several technologies and concepts, which are intertwined together in a mind-boggling way; see \cite{deng2020edge}.

Deng et al. \cite{deng2020edge} stated that there has not been a formal and globally accepted definition of edge intelligence today. However, the definition is given in some studies. To be specific, Xu et al. \cite{xu2020edge} defined edge intelligence as a new paradigm of intelligence involving a collection of connected systems and UEs for collecting, analyzing, processing, and caching data near the sources of the data. The goal is to improve the data processing speed and quality and to secure and protect data privacy. Hu et al. \cite{Corici2018Edge} described edge intelligence as the paradigm shift involving data collection, transmission, processing, and caching through the use of edge computing with ML techniques and higher networking capabilities. 

Contrasted with cloud-based intelligence, in edge intelligence, data is locally analyzed and processed. In edge intelligence, a distributed computing paradigm offers edge inference, edge training, edge caching, and computing services at other edge devices or edge servers for the requirements of a particular edge intelligence application. Thus, edge intelligence brings striking gains by effectively protecting the subscribers' privacy, saving bandwidth, ensuring higher reliability, and lessening response time. Furthermore, by training ML and/or DL models with self-created data, edge intelligence allows subscribers to customize smart applications. In intelligent edge, AI offers strong support for edge computing. The focus of intelligent edge lies in solving edge computing problems with the AI techniques, such as resource allocation optimization \cite{xu2020edge}. Both intelligent edge and edge intelligence require each other. In fact, the DL services in intelligent edge are likewise a piece of edge intelligence. Hence, in addition to resource utilization, intelligent edge can offer enhanced service throughput for edge intelligence \cite{wang2020convergence}.

\section{COMMUNICATION, COMPUTATION, CACHING, AND CONTROL MODELS}

A model may represent a theory and plays a crucial role in simplifying the real-life situation analysis. In this section, we explore four different models of 4C to lay the foundation of resource integration for 5G, 6G, and beyond networks. Thus, beneath this section, we discuss the communication model in Section IV-1, the computation model in Section IV-2, the caching model in Section IV-3, and the control model in Section IV-4. As per \cite{mao2017survey}, such models can support mechanisms for abstracting diverse functions and operations into optimization problems and simplifying theoretical analysis.

\subsection{Communication Model}

In recent decades, stochastic geometry was introduced to serve as a standard tool for modeling and designing wireless networks. A rich set of spatial point processes comprising Poisson Point Process (PPP) and cluster processes have been employed for modeling node locations in various wireless networks, including Het-Nets, cellular networks, and cognitive radio networks. Many research efforts in this area were devoted to addressing interference and wireless channels hostility, such as fading and path loss, to guarantee higher coverage and channel reliability for RAN or distributed D2D networks \cite{ko2018wireless}. Actually, improving the networks' performance in terms of throughput, low latency, and spectral/energy efficiency has been the main emphasis throughout the evolution of mobile communication networks \cite{liu2016three}. Hence, several parameters need more attention for efficient wireless networks design and network resources optimization.

\subsubsection{Spectral Efficiency (SE)}

Various wireless networks air interface techniques, including adaptive modulation and coding (AMC), multiple-input-multiple-output (MIMO) antenna strategies, and frequency domain packet scheduling (FDPS), have improved SE to a great extent. Today, such techniques extend SE near the theoretical Shannon's capacity limit \cite{baumgarten2014lte}. However, Shannon's theory remains a key design base for the emerging 6G wireless network and offers two main approaches of maximizing network capacity: i) increasing network bandwidth and ii) improving SE. As a key performance indicator (KPI) for the 6G wireless network design and analysis, SE has to be further improved to tackle issues of communication resources, such as multi-dimensional radio and x-haul resources. Thus, in 5G and 6G networks, SE will be enhanced to 3$\times$ that of 4G and 5-10$\times$ that of 5G, respectively \cite{zhang20196g}. As per \cite{jiang2021road}, 30\textit{ bps/Hz and 15 bps/Hz} in the downlink and uplink, respectively, mark the minimum requirements for peak SE in 5G. Realizing this will enable efficient computation offloading, which serves a significant role in the i4C. Based on \cite{tan2017virtual}, we can express the achievable SE for offloading tasks in an uplink direction in (1) as

\begin{equation}
	\ell_{u}^{k}=\log _{2}\left(1+\frac{p_{u}^{k} G_{u}^{k}}{\sigma^{2}}\right),
\end{equation}
where  $p_{u}^{k}$ denotes the transmission power from UE to an SBS $k$ and the $G_{u}^{k}$ represents the corresponding channel gain between UE $u$ and SBS $k$.

\subsubsection{Interference}

Now that the state-of-the-art wireless technologies operate nearer the Shannon capacity bound, a limited capacity gain can be extracted with current cell structures and frequency allocation techniques. The 5G radio access technology (RAT) promises to utilize a three-dimensional capacity model (i.e., bits per second $\times$ Hertz $\times$ cells per square kilometer), implying that the so-called capacity gain can be achieved with an increase in a number of cells per square kilometer. Cell densification has been earmarked to generate more capacity gain for 5G networks \cite{saha2017evolutionary}. 6G is expected to be more heterogeneous than 5G, thus presenting more promising scenes for computation offloading due to the proximity of UEs to SBSs, higher capacity, and lower latency. However, the increase in SBSs raises high energy costs, and the closeness of SBSs to one another can generate severe co-channel interference \cite{wang2016power,zhang2016regularized}. Co-channel interference can immensely complicate the computation offloading decision, which is determined by the wireless transmission condition. Hence, the interference mitigation techniques, such as transmission power control, frequency subcarrier allocations \cite{mu2019latency}, adaptive beamforming, interference cancelation, interference randomization \cite{hamza2013survey}, and coexisting cloud and edge AI \cite{zhang20196g} should be considered in the future wireless networks design to improve the rate of offloading tasks.

\subsubsection{Bandwidth and Power Allocation}

Bandwidth and power allocation has been a pivotal technique of improving network efficiency under guaranteed QoS to users. In orthogonal frequency division multiplexing (OFDM), bandwidth is allocated by converting a wideband spectrum into several narrowband orthogonal subcarrier channels to serve multiple users at a time. That means the subcarrier channels can be shared among several users using the same network concurrently; hence, users can efficiently offload the tasks via the subcarrier channels. Moreover, in multicarrier systems, the total transmitted power (the power required on each subcarrier) is minimized to control co-channel interference while offloading the tasks for computing and caching services. Such technique dramatically reduces interference since a subcarrier can be occupied by at most one user \cite{sharnagat2015method}. Thus, the demands for higher throughput, lower latency, and higher reliability will be met. On top of that, 6G holds strong potential to emerge with superior wireless channels, such as universal filtered multicarrier and filtered-OFDM, which could further accelerate the computation offloading efficiency.

\subsubsection{Energy Efficiency}

The increasing demands for higher communication capacity and fast-growing energy costs make the energy-efficient wireless communication network design an emerging trend. In conventional networks, the radio access part is viewed as the prime energy consumer, accounting for greater than 70\% of the total energy consumption \cite{zhang2017quantcloud}. In the design and analysis of green networks, EE metrics are intrinsic since they help assess and compare the consumed energy of various designs and provide long-term research goals \cite{yang2018characterizing}. Designing an energy-efficient wireless network is highly desirable for efficient tasks offloading. Hence, EE has to be improved in the future wireless networks design. According to \cite{zhang20196g}, the network EE will enhance to 10-100 $\times$ that of 4G and 10-100 $\times$ that of 5G, in 5G and 6G networks, respectively. Therefore, as a KPI for evaluating 6G wireless networks, EE has to be thoroughly considered.

\subsubsection{Achievable Transmission Rate}

Several studies explored the maximum limits of the achievable transmission rates of wireless communication networks over fading channels \cite{deng2004information}. The study of time-varying fading channels, where fading gains (channel states) are often modeled as stochastic processes, such as independent and identically distributed (i.i.d.) process and Markov process, yield tremendous achievements \cite{liu2015capacity}. For modeling wireless fading channels, finite-state Markov channel (FSMC) is mostly considered. The FSMC model, which relies on partitioning the received signal-to-noise ratio (SNR) into a finite number of states, has drawn much attention due to its great balance between complexity and accuracy \cite{wang2008new}.

The efforts in \cite{wang2019edge,he2018integrated,he2018trust} modeled the wireless channels as FSMC; the goal is to effect higher efficiency than that of traditional assumption of static channels. In such wireless scenario, Shannon's theorem can be used to evaluate the achievable data rate. In other words, in a cellular cell with an available bandwidth allocated to each UE and a given BS transmission power, the ultimate wireless transmission rate of a mobile UE is defined by the Shannon's capacity theorem \cite{luo2017energy}. Hence, based on the achievable SE $\ell_{u}^{k}$ expressed in (1), the achievable transmission rate of the UE $u$ can be expressed as;

\begin{equation}
	R_{u}^{k}=\psi_{u}^{k} B \ell_{u}^{k},
\end{equation}
where $B$ represents the available communication bandwidth and $\psi_{u}^{k}$ denotes the fraction indicator (0 $\leq$ $\psi_{u}^{k}$ $\leq$1) of the available bandwidth allocated to the UE $u$.

The time (in seconds) taken to offload a computational task $n$ from a mobile UE $u$ through a wireless channel, i.e., transmission time/delay can be obtained by dividing the task input-data size (denoted by $L_n$) by the achievable transmission rate expressed in (2). Hence, transmission time does not depend on the channel length, rather, it relies on the input-data size/quantity of the task and the data transmission rate, and it can be given by;

\begin{equation}
	T_{u,n}^k = \frac{L_n}{R_u^k},
\end{equation}
where $L_n$ (in terms of bit) represents the input quantity of the task $n$; see \cite{tan2017virtual}, \cite{cheng2018energy}.

As per \cite{cheng2018energy}, the energy (in joule) consumed while transmitting a task $n$ from a mobile UE through a wireless channel can be obtained simply by multiplying the transmission power of the UE $u$ ( $p_u^k$) and the transmission time expressed in (3). This energy can be expressed as;

\begin{equation}
	E_{u,n}^{k}=\psi_{u}^{k} p_u^k\frac{L_n}{R_u^k}.
\end{equation}

In wireless networks, the uplink/downlink transmission of UE is generally managed by a wireless BS, which might be a Wi-Fi AP, a Femtocell network AP or even a macro-cell BS \cite{chen2014decentralized}. RAN is widely seen as an intrinsic part of the wireless network infrastructure facilitating wireless connection between the UE or any wireless controlled device and the cellular core network. In an MEC system, the UE is wirelessly connected to RAN; likewise, RAN is connected to the core network through the guided channels like IP/ethernet. In particular, RAN provides connection between BSs and backhaul networks through the ethernet interface, supporting high-speed data transmission \cite{abbas2017mobile}.

In mobile cloud computing (MCC), the communication paths between the UEs and cloud servers are normally abstracted as bit pipes characterized by constant or varying rates with given distributions. The point behind such models has to do with tractability and may be connected with an MCC system design, focusing on handling the latency issues in the core network and managing the large-scale cloud. Thus, these bit-pipe channel models do not pay attention to the wireless communication network latency. Conversely, in a small-scale edge cloud, such as an MEC system, the focus is to dramatically minimize the communication network latency through an advanced air interface design. Moreover, the bit-pipe models omit some important features of wireless propagation and are too simple to support implementing modern communication techniques \cite{mao2017survey}.

\subsection{Computation Model}

\subsubsection{An Overview}

Computation offloading associates with communications through two major issues, i.e., latency and energy consumption. Nonetheless, it plays a pivotal role in computation aspects \cite{barbarossa2014communicating}. In literature, several offloading destinations were considered for executing the computational tasks. The main destinations include: i) cloud-only, in this class, the computation-intensive tasks and latency-tolerant applications are pushed to the remote cloud server for execution, ii) MEC-only, herein, the latency-sensitive applications, e.g., VR, Tactile Internet, face recognition, and other mission-critical IoT applications, are migrated onto an MEC server for speedy execution, and iii) local, where the tasks are rather executed locally at UEs than at an MEC/cloud server due to higher energy consumption and latency \cite{rimal2016mobile}. In \cite{singh2017optimize}, three key steps were identified for offloading computation-intensive tasks from UEs; these steps are: i) tasks partitioning, ii) offloading preparation, and iii) offloading decision, as shown in the sequel.

\textit{\textbf{Tasks Partitioning:}} The tasks here are grouped into offloadable and non-offloadable components (tasks). The former involves the tasks that should be migrated for remote execution; whereas the latter refers to the tasks that should be retained at UEs. Identifying computation-intensive tasks to be offloaded requires performing source code analysis and performance prediction by an application programmer \cite{singh2017optimize}. Determining part of the tasks that should be offloaded for execution is important in partial offloading; thus, there is a need for partitioning tasks into modules. The partitioning approach can either be static (where the tasks are partitioned into a fixed number of partitions during application development) or dynamic (where the tasks are partitioned at runtime based on the availability of bandwidth and quality of network connection). The static is quite easier to implement. However, the dynamic is important in cases where the computation-intensive applications are required to adapt to the network and mobile environment changes. For instance, some frameworks, such as MoSeC and Self Cloning, employ dynamic partitioning; whereas others like Aura, Avatar, and MALMOS use both static and dynamic \cite{deshmukh2016computation}.

\begin{figure}[t!]
	\includegraphics[width=9cm, height=6cm]{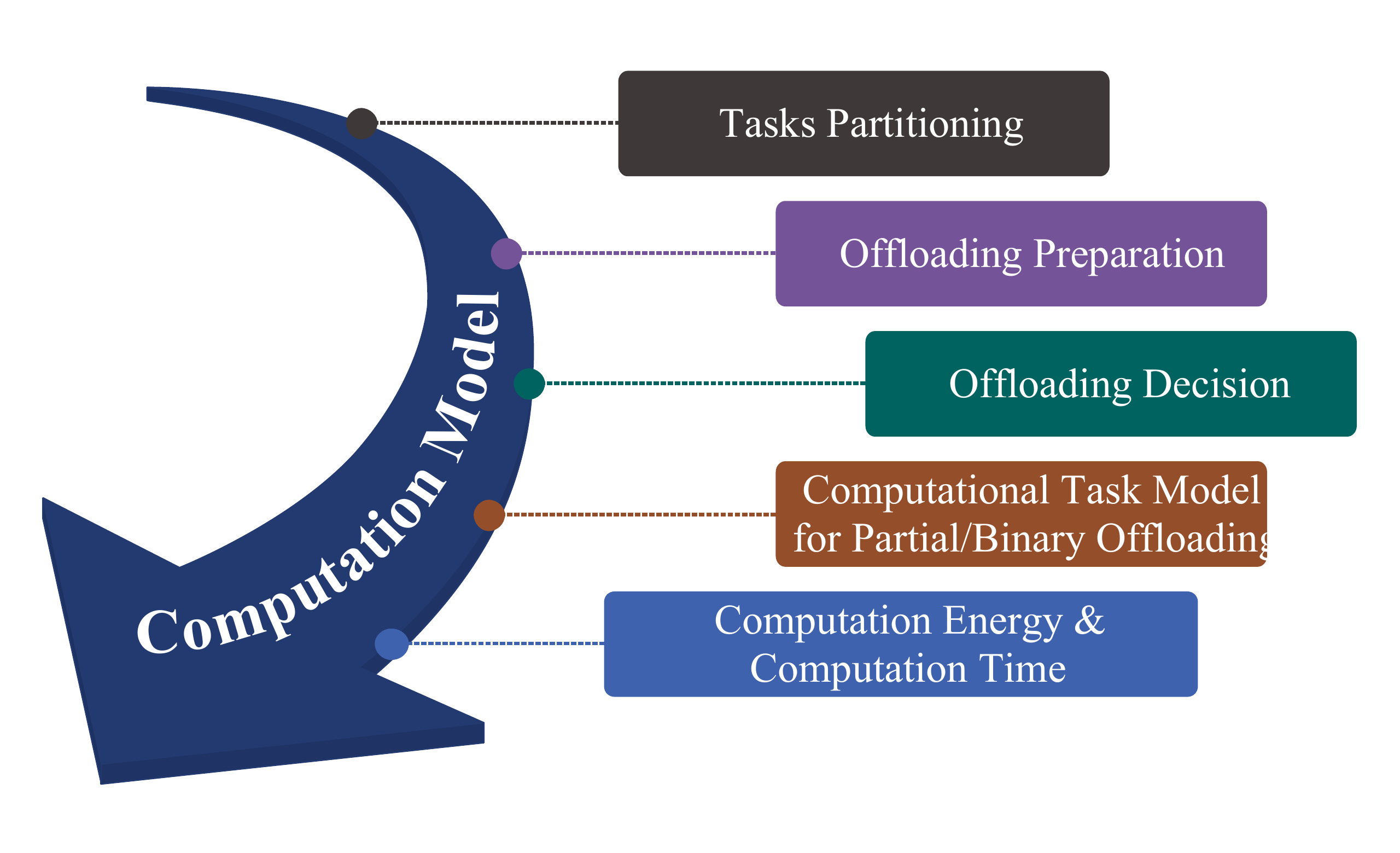}
	\caption{A rodmap of the computation model.}
	\label{fig6}
\end{figure}

\textit{\textbf{Offloading Preparation:}} This covers all essential steps, such as remote server selection, migration, code installation, and tasks or data migration for remote computing, required for offloadable tasks to enable their use in native UE applications \cite{singh2017optimize}.

\textit{\textbf{Offloading Decision:}} This usually precedes remote execution of offloadable tasks. The execution context determines whether an installed remote task is used in the UE applications or not \cite{singh2017optimize}. Offloading decision is an extremely complex process affected by various factors, including subscriber's preferences, application (nature), UE capability (e.g., high or low performance UE), application model, connection (e.g., network bandwidth, delay, and costs), and cloud/MEC service \cite{othman2013survey,mach2017mobile}. Generally, offloading decision is carried out in sequence. This can be explained in an MEC environment, in such scenario, UE decides whether to offload a computational task via wireless channels for remote execution or execute them locally. If the task is offloaded to the network edge/MEC server, the server determines whether it can meet the request or the computation should be further offloaded to the remote cloud for execution. Offloading decision can either be full or partial. The gains derived from full offloading decision are to: reduce the consumed energy at UEs while satisfying the constraints of delay, ii) lower the computing delay, and iii) determine a decent tradeoff between the computing delay and energy consumption. In partial offloading, the benefits are to: lower the consumed energy at UEs while satisfying the constraints of computing delay and ii) find appropriate tradeoff between the computing delay and energy consumption \cite{mach2017mobile}.

Before walking through the computation model, we find it helpful to briefly look into partial offloading and binary offloading at this point. The former enables partitioning of computational tasks into different parts at UEs for local computing and offloading simultaneously. Whereas the latter makes it impossible to partition the computational tasks; rather, the tasks are executed as a whole either at the UEs or at the network edges \cite{liu2017price}. To simplify the description of the remote CPU computation model, we restrict the scope of our discussion to the computational task model for binary offloading, which is subsequently followed by computation energy and computation time as following.

\subsubsection{Computational Task Model for Binary Offloading}

Here, the task model can be described in terms of two or three different fields. In a two-field task model, a computational task $n$ can be characterized by $T(L_n$, $D_n$). In this case, $L_n$ $\textgreater$ 0  (in bits) and $D_n$ $\textgreater$ 0, respectively, depict the task input size and the total number of the required CPU cycle to execute the task \cite{chen2014decentralized}. The size of the computation output is usually assumed negligible; thus, it was ignored in \cite{chen2014decentralized}, \cite{cao2018joint,mao2016dynamic,sheng2015energy}. There are various factors on which the required CPU cycle number depends for executing the tasks. These factors include specific applications, task input size, and physical components, such as memory and CPU, in the computing device \cite{cao2018joint}. However, the number of the required CPU cycle is not considered in some two-field notation models; rather, a task completion deadline might be a priority especially when computation delay becomes a concern. Such model was adopted in \cite{cao2018joint,mao2016dynamic,sheng2015energy}, where $T$($L_n$, $\tau_n$)) characterizes the computational task $n$. This implies that each task $n$ with an input-data size $L_n$ has to be accomplished within its completion deadline, denoted by $\tau_n$ $\textgreater$ 0 duration. For the latency-sensitive applications, it is assumed that the completion time cannot exceed the coherence time of the channel. Thus, the channel power gain does not change within the block of interest.

The computational task requested by the UE applications can be modeled as i.i.d. Bernoulli process. To be specific, for each time slot, the probability of requesting a task can be denoted by $\rho$, and that of not requesting it can be given by 1-$\rho$. Such request is often followed by an execution decision \cite{mao2016dynamic}. For instance, when a mobile VR device requests a computational tasks, an MEC server determines whether the requested tasks should be computed on it or not. If the tasks should not be executed at the server, the tasks or their corresponding parts (chunks) have to be forwarded to the VR device for local execution \cite{yang2018communication}.

Alternatively, the three-field task model has been adopted in several efforts, including \cite{mao2017survey}, \cite{cheng2018energy,guo2017energy}. In such model, a three-field notation $T$($L_n$, $X_n$, $\tau_n$) can be applied to represent a computational task $n$. This notation carries the information of the task input size $L_n$ $\textgreater$ 0 (in bits), the computation workload/intensity $X_n$ $\textgreater$ 0 (in CPU cycles per bit), and the task completion deadline $\tau_n$ $\textgreater$ 0 (in seconds) \cite{cheng2018energy,yang2018communication,guo2017energy}. The three fundamental parameters of a task are determined by the nature of the task itself \cite{mao2017survey,cheng2018energy,guo2017energy,cui2017energy}.

The CPU workload of a computational task directly determines the consumed energy of computing. The workload is determined by the total amount of the required CPU cycles. In short, the relationship between the number of the required CPU cycle ($D_n$) for computing a task $n$ and the task input size ($L_n$) with the computation intensity ($X_n$) can be expressed as: $D_n$= $L_n$$X_n$. Hence, the number of the required CPU cycles for executing computational tasks differs in various applications, and it may be determined by offline measurement \cite{mao2016dynamic}.

\subsubsection{Computation Energy}

The power consumption in a CPU depends on a number of factors, such as short circuit power and dynamic power. However, the main energy consumer is dynamic power, which can be controlled by adjusting the CPU-clock frequency of the chip voltage based on the dynamic voltage and frequency scaling (DVFS) mechanisms \cite{sheng2015energy}. The DVFS varies the CPU supply voltage and clock frequency of UEs according to computation load to meet the performance requirements. With the DVFS mechanisms, the CPU-clock frequency of UEs will be regulated to lower the energy consumption in an adaptive way. Therefore, by incorporating DVFS techniques into computation offloading, the strategy design becomes more flexible \cite{zhang2018energy}.

For the computational task $n$, the total energy consumption constitutes the energy consumed while offloading the task from the UE to the network edge (e.g., SBS $k$) and the energy consumed while computing the task at the SBS. This is given by$E_{u,n}^k$ (as expressed in Section IV-1) and $E_{u,n}^k$, where $E_{n}^k$=  $\mu_0$$f_{k}^2$$D_n$, $\mu_0$ is a constant related to the CPU of a computing server in the SBS $k$, and $f_k$ denotes the CPU cycle frequency of a computing server in the SBS $k$. Here, the computation result is assumed negligible \cite{cheng2018energy}. To that effect, the total energy (in joule) expended for executing the task $n$ at the network edge can be represented as

\begin{equation}
	E_t=E_{u,n}^k + E_{n}^k.
\end{equation}

\subsubsection{Computation Time}

Here, our focus lies in the total time costs for executing a specific task at a network edge, consisting the time consumed while transmitting the task to the network edge and the actual time spent while executing the task at the edge. Thus, a computational task $n$, denoted by $T$($L_n$, $X_n$, $\tau_n$), can be offloaded through an OFDMA channel with maximum achievable transmission rate to the network edge (e.g., SBS $k$). In this regard, the time expended for offloading the task $n$ to the SBS $k$ can be given by $T_{u,n}^k$, as expressed in Section IV-1. Likewise, the time cost for executing a task at the network edge, as defined in \cite{cheng2018energy}, refers to the total amount of the required CPU cycles for computing the task divided by the corresponding CPU-cycle frequency. Hence, the time consumed for computing the task at the SBS $k$, which depends on both CPU-cycle frequency $f_k$ and computation intensity, can be given by $T_n^k$ = $\frac{L_n X_n}{f_k}$ ; see \cite{cheng2018energy,guo2018energy,chen2015efficient}. To that effect, the total time cost (in second) for executing the task $n$ at the network edge can be expressed as

\begin{equation}
	T_t=T_{u,n}^k + T_{n}^k.
\end{equation}

\subsection{Caching Model}

\subsubsection{An Overview}

In the previous decades, research studies devoted to CDNs gained momentum by focusing on where to deploy the servers (server placement); which files to cache at each server (content placement); how much storage capacity to allocate to each server (cache dimensioning); and how to route content from caches to end-users (routing policy) \cite{paschos2018role}.  

The caching systems may differ in terms of granularity, scale, and technologies. Nonetheless, the common goal shared by all caching systems is to optimally cache data contents for subsequent usage \cite{tarnoi2019adaptive}. To provide clear description of the caching model, this overview touches on the two basic approaches to caching studies, caching places, and performance of cache networks.

\textit{\textbf{Approaches to Caching Studies:}} The effort in \cite{shivaram2018queuing} described two basic approaches to caching studies, i.e., coded and un-coded caching. One of the approaches involves the conventional caching schemes, such as first-in-first-out (FIFO), least frequently used (LFU), least recently used (LRU), etc. Such schemes generally use cache hit ratio as a key parameter for performance evaluation and are often called un-coded caching schemes because there is no coding in them. Besides, each of the schemes is featured with particular insertion and eviction policies. The other approach, i.e., coded caching, involves content placement and content delivery. In the placement phase, caches are populated with file contents usually during low network activity, i.e., off-peak hours. While in the delivery phase, the server serves the requests by executing coded multicasting. The peak/average number of file contents transmissions via the shared link is commonly considered as a key performance metric. Since coding is used for delivering content, this approach minimizes the file transmissions within the networks and attracts extensive research investigations; see \cite{maddah2014fundamental,maddah2016coding}.

\textit{\textbf{Caching Places:}} In mobile networks, several places can be considered for deploying edge servers and content caching. The three key places where cache can be deployed to cache file contents in cellular networks constitute UEs, RAN, and the core network \cite{wang2017survey,safavat2020recent}. Since deploying cache at the evolved packet core (EPC) is technically more convenient than deploying cache at RAN, EPC is considered as the most commonly deployed caching place. At the network edges, content may be cached in MBS, SBS, or UEs \cite{wang2017survey}. In MEC, caching content at the network edge brings significant gains, enabling MEC to get real-time information from RAN and utilize it for guaranteeing QoE of UEs. Hence, with real-time information, remote/MEC server optimizes the users' traffic to ensure QoE \cite{tan2018radio}.

\textit{\textbf{Performance of Cache Networks:}} The main considerations for overall caching performance involve caching policies, deciding what to cache and when to deliver the caches \cite{wang2014cache}. Hence, user's request patterns, caching policies, and how caches are operated (cooperatively, independently, or in a globally coordinated manner) constitute some of the several factors on which the performance of cache networks depends. The caching policy is vital; upon a user's request for data, it makes decisions whether to cache the data, where to cache it, what timer value to set in case of using time-to-live caches, or which data to evict in case of a full cache \cite{ramadan2019performance}. Various caching policies have been offered in literature for managing a single cache, which differ in terms of either eviction or insertion policy. In \cite{martina2014unified}, some existing caching policies, such as FIFO, LFU, LRU, q-LRU, k-LRU, RANDOM, and k-RANDOM were studied. 

Being easy to implement, LRU is widely adopted and it provides great performance. In the context of ICN, FIFO and RANDOM are considered as feasible substitute for LRU since their hardware implementation in speedy routers is easier. The k-LRU and q-LRU enhance the LRU performance through advanced insertion policy; see \cite{martina2014unified}. Estimating the gain behind a content by assessing its present popularity, potential popularity, cache capacity, and locations of existing replicas over the network topology is essential. Instead of applying conventional policies, including FIFO, LFU, and LRU, it is desirable to propose cooperative caching policies for EPC and RAN caching in order to efficiently increase cache hit rate \cite{wang2014cache}.

The concept of caching has been advancing rapidly due to the unprecedented growth of network traffic over wireless networks. With the advent of MEC, both caching \cite{mehamel2018energy} and computing are pushed near the UEs, i.e., network edges. This implies dramatic reduction in both content delivery delay and network traffic and also guarantees adequate computing and caching functionalities. Indeed, efficient caching functionality at the edges allows mobile UEs to alleviate the possible burden from backhaul links. Thus, caches can be designed for efficient communication between edge servers and UEs \cite{tan2018radio}, \cite{mehamel2018energy}. Moreover, the classification of caching techniques, not only in the context of MEC, can be either reactive/transparent or proactive. In transparent caching, neither UE nor the application service provider (ASP) is aware of a caching MEC server. In proactive caching, data contents are non-transparently cached before being requested since it can result in high network utilization in future \cite{beck2014mobile}.

\subsubsection{Cache Performance Metrics}

In conventional caching scheme, cache-hit rate/cache-hit ratio is considered as a common performance metric, representing the ratio of the requests satisfied by a caching system and the aggregate incoming requests. Generally, high cache-hit ratio implies a high-performance caching system since it brings about dramatic reduction in redundant data transit. There are four factors on which cache-hit ratio depends, i.e., cache size/capacity and cache algorithm; these two can be figured and controlled to a certain extent. The other two factors are content population and content popularity distribution; these cannot be controlled. They are externally generated by subscribers and corresponding applications that interact with the caching system. Despite being widely viewed as a common performance metric of caching systems, cache-hit ratio is incapable of providing insights into the performance of network of caches. Thus, being a conclusive result of request filtration that all caching systems in a network achieve, server hit ratio is considered as a more appropriate metric. Another KPI is footprint distance. Here, the shorter the footprint distance, the nearer the data content is to the requesters, hence implying shorter response time for the requesters. In short, for green communication networks, caching time is an essential index. Hence, in designing a cache algorithm, footprint distance, server hit ratio, and caching time have to be considered \cite{tarnoi2019adaptive}.

\subsubsection{Cache Capacity}

The cache capacity or size of cached information (in bytes) is widely considered as the typical measurement of caching capability \cite{wang2017integration,liu2016three}. With increase in the capacity of cache memory, the cache-hit ratio can be improved. Compared with content population, the cache capacity is relatively small. However, multi-magnitude increase in the cache capacity may lead to a few percentage points of cache-hit ratio improvement \cite{shivaram2018queuing}. At the network edge, after computing the tasks, the outputs might be considered reusable. In such case, a BS with a given caching capacity, say $C$ bytes, can be required to cache the outputs \cite{cui2017energy}. Therefore, to cache content at the network edge, each BS decides whether the content offloaded from UEs should be cached in its cache memory before or after computation based on each content's popularity distribution. This is achieved by considering two binary parameters to control the caching strategy, e.g., $\gamma_u^{k,1}$ and $\gamma_u^{k,2}$. If it is decided that BS will cache the original content $\gamma_u^{k,2}$ is set to 1; otherwise, it is set to 0. On the other hand, if it is decided the BS will cache the computed content $\gamma_u^{k,2}$ is set to 1; otherwise, it is set to 0 \cite{zhou2017information}. The cached content may be represented by a set of finite numbers, where each content can be a short video clip or a portion of movie with a given size (in bits) \cite{kwak2018hybrid}.

\subsubsection{Cache Capacity Constraint}

Since the capacity of cache memory of a BS (e.g., MEC server) is finite, the total size of cached content cannot exceed it \cite{zhou2017information}. That is to say the caching capacity constraint, expressed in (7), has to be satisfied. Specifically, for a given cached content, the caching capacity constraint can be expressed as

\begin{equation}
	\sum \gamma_{u}^{k,2}L_r \le C,
\end{equation}
where $C$ (in bytes) denotes a given caching capacity and the cache decision variable and computation result can be denoted by $\gamma_u^{k,2}\in$ $\{0,1\}$ and $L_r$, respectively \cite{cui2017energy,wang2017computation}.

In the conventional (un-coded) caching scheme, the gain is derived from making content available locally. Actually, a UE may request some content cached in its cache; in this way, the local memory of the UE serves this request. This implies the local caching gain, which is essential if the local cache memory is big enough to cache ports of the popular content locally. In the coded scheme, as per \cite{yang2018communication,guo2017energy}, the global gain is derived from joint optimization of the placement phase and delivery phase, ensuring that various requirements are met in the delivery phase with the single coded multicast transmission. Because content placement is carried out without knowing the actual requirements, realizing the global gain requires careful design of the placement phase such that multicasting opportunities can be created at the same time for all possible requests in the delivery phase. In a word, if the aggregate global cache capacity exceeds the total content size, then the global caching gain becomes relevant \cite{bouras2019synergy,maddah2014fundamental,maddah2016coding}.

\subsubsection{Caching Reward}

In mobile wireless networks, the reduction of the network backhaul delay or the backhaul bandwidth alleviation is considered as a caching reward. Hence, the reward of caching content requested by a mobile UE can be expressed by $\gamma_u^{k,2}Rc$. In this regard, $c$ and $R$, respectively, represent the content request rate requested by a mobile UE u and the average data rate of a single UE in the system \cite{wang2017computation}.

\subsection{Control Model}

\subsubsection{Distributed Control Model}

Designing an accurate control model that controls, coordinates, integrates, and optimizes communications, computing, and caching resources can be highly complex. Recently, \cite{ndikumana2019joint} adopted a distributed control model based on a distributed optimization by which the communication, computing, and caching models can be coordinated and integrated at the network edges. The distributed control model enables the MEC servers hosting computing and caching resources to be deployed at the same domain and collaborate to share resources. In this way, the information exchange between the MEC servers and centralized cloud server can be reduced, thereby minimizing the backhaul bandwidth. Hence, with the distributed control model, the cache hits can be improved. As discussed in Section IV-3, the reduction of backhaul delay is termed caching reward. Here, the amount of backhaul bandwidth saved by the distributed control is adopted as a caching reward.

Moreover, the distributed control enables the exchange of a small amount of information among the collaborating MEC servers, thus bringing significant gain in terms of maintaining the resource allocation in the vicinity of accessible computing and caching resources. However, among the collaborating MEC servers, there is no provision for a centralized controller that controls the entire servers. In such a case, the distributed control can be modelled as dynamic feedback control model; see \cite{farivar2015local}. So then, the resource allocation table update at each server can serve as a feedback with a given state at iteration t, which is used for determining the new state at the next iteration $t + 1$; see \cite{ndikumana2019joint}.

Therefore, instead of collecting all problem parameters and performing a central calculation, several agents, obtaining certain problem parameters by sharing information with finite set of neighbors, compute distributed algorithms. The agents may represent BSs, edge/MEC servers, UEs, or buses depending on the specifics of the distributed algorithm and the application of interest. Compared with centralized approaches, distributed control algorithm brings several gains. For instance, the computing agents can only exchange small amount of information with a subset of the other agents. This means the expense of the challenging communication infrastructure can be lowered and also cybersecurity can be improved. Due to its ability to do parallel computations, distributed control algorithm can be computationally superior to centralized control algorithm when it comes to the maximum problem size that can be handled and solution speed. In addition, distributed algorithm is robust in terms of failure of individual agents. Finally, distributed algorithm has the potential to respect data privacy, cost functions, measurements, and constraints, which becomes more significant in a distributed generation scenario \cite{molzahn2017survey}.

\subsubsection{Hierarchical Control Model}

In hierarchical modelling, models may represent different parts of a studied system or its various properties that are logically ordered to form a hierarchy or a sequence. In modelled systems description, the lower hierarchical levels usually correspond to higher levels of detail. Besides, there is nearly similar level of detail in each element of a sequence, and the outputs of a present model imply the input data of a succeeding/next model \cite{novikov2016hierarchical}. Molzahn et al. \cite{molzahn2017survey} described hierarchical control scheme as algorithms where computations are performed by agents that exchange information with other agents at a higher level in a hierarchical structure, eventually leading to a centralized control. The promising gains of hierarchical control triggers extensive research efforts.

In \cite{kim20185g}, hierarchical control scheme was proposed to model the interactions between SBSs and UEs by integrating bandwidth allocation for communication, computation offloading, and cache splitting. The control scheme harnesses the synergistic combination of caching, computing, and communication capabilities in the SBS, hence characterizing competitive and collaborative interactions among them. Specifically, a two-tier hierarchical game model was applied based on a unified and integrated approach to model the interplay between the SBSs and UEs. The control decisions are made by the game players, i.e., SBSs and UEs, in line with the step-by-step timed learning approach. In the first-tier, only SBSs act as the game players. The communication bandwidth is shared among these SBSs as per a dynamic bargaining model. In the second-tier, each SBS and its respective UE act as the game players, and the interactions among them are modelled as Stackelberg game model. The SBS, serving as a leader, splits its caching capacity and decides the cost of communication and computing services. The leader's decision is monitored by the UEs (followers), which select their appropriate strategy. In \cite{ndikumana2019joint}, the hierarchical control enables offloading decision-making at UEs and enables each MEC server, as a controller, to decide for the offloaded tasks.  In a word, control entails decision-making with respect to what, how much, where, and when to allocate available resources to a given task.

 However, limited resources and heterogeneous users’ demands make resource allocation a complex problem. Generally, the types and amount of the required resources are determined by users in their requests. Network service providers, in response, allocate the resources requested while ensuring that the granted resources are sufficient to meet the constraints defined by the users. As a result, any control strategy for resource allocation systems i) has to be focused on requested resources, ii) must satisfy the users’ ever-changing demands, iii) has to be optimal by ensuring optimal utilization of resources, and iv) prioritize task for superior performance \cite{abid2020challenges}. These requirements make resource allocation system control very challenging and thus limited. Hence, coming up with tractable control techniques that guarantee optimal distribution/allocation of a system’s resources is not an easy task. Nearly all the optimal or near-optimal resource allocation system control techniques/algorithms are not tractable. On this account, researchers make a trade-off between tractability and optimality. Recently, Lima et al. \cite{lima2020model} achieved near optimality of control system and stability by emplying the constrained reinforcement learning (RL) techniques. This achievement portrays strong potential of intelligent control in overcoming the shortcomings of conventional resource allocation control techniques.

\subsubsection{Intelligent Control Model}

This involves learning, decision-making, and optimization based on the frontiers of AI, e.g., DL, ML, and DRL. Intelligent control relies on utilizing existing knowledge or experience to enable various agents to intelligently learn, optimize, and take appropriate actions (e.g., resource allocation control, network association, and resource management) with dual functions for supporting diversified network services. Such functions can be realized with AI techniques applied in 6G networks. Thus, network agents, such as BSs, MEC/edge servers, and UEs, can be equipped with learning models (intelligent brain) such that they automatically learn to make resource allocation decisions \cite{yang2020artificial}. Generally, edge devices or edge servers host both edge resources and training at the network edges. Such servers are not powerful as computing clusters or centralized cloud servers. Xu et al. \cite{xu2020edge} raised four major problems that need consideration for edge training, i.e., i) how to train (the training architecture), ii) how to speed-up the training (acceleration), iii) how to optimize the training approach (optimization), and iv) how to assess the vulnerability of the model outputs (uncertainty estimates).

Today, the intelligent control brings striking gains in terms of coordinating, controlling, and optimizing mobile communication, computing, and caching resources. Considering the heterogeneous nature of both wireless networks and UEs with challenging QoS requirements of the UEs applications, the conventional resource allocation optimization and control algorithms cannot be sustainable for performance requirements of 6G and beyond mobile networks. Hence, the promising answer lies in the AI-based control (which can be distributed algorithms/model), which has been attracting extensive studies.

To realize the i4C framework, these models can be jointly optimized by considering the possible constraints, decision variables, and optimization objectives. Depending on the specifics of the optimization and the application task of interest, i) the decision variables comprise the computation offloading, data caching, execution, and resource allocation decision variables; ii) the optimization goals could be network performance, costs of deployment and operation (communication/computation costs, energy consumption, etc.), efficiency, network/system reliability, and privacy \cite{deng2020edge}; and iii) the constraints may include computation time, local computing capabilities of UEs, caching and computing capacities \cite{ndikumana2019joint}, wireless channel states, dynamic trust values, cache status \cite{he2018integrated}, computation delay \cite{cao2018joint}, and so on. Thus, various optimization techniques can be considered to realize the i4C model.

 Note that, some traditional optimization techniques may not yield desirable results due to the numerous configurable parameters, wireless channels conditions, multiple decision-making variables, and heterogeneous users’ demands; besides, the complexity of mobile networks is still growing. However, among the conventional optimization approaches, the Lyapunov optimization method excels as the best candidate for the long-term stability of dynamic systems. What follows in the next section is devoted to various optimization approaches for achieving the i4C in 6G and beyond networks.

\section{INTEGRATION OF 4C}

This section presents a great deal of research efforts aimed at combining 4C. We observe that several cutting-edge research efforts on the i4C focused on applying the AI techniques to integrate/optimize 4C at the network edges due to the growing complexity of mobile wireless networks. This is in contrast to earlier efforts that tended to integrate/optimize the resources based on the conventional resource allocation optimization approach. This section considers both conventional and recent integration approaches. Specifically, the section reviews recent works on the i4C based on the conventional optimization approach in Section V-1, discusses intelligence for resources integration in Section V-2, reviews recent trends in AI-based integration approach in Section V-3, brings a snapshot of the various approaches devoted to the integration of 4C in Section V-4, and discusses the integration of sensing and communication in Section V-5.

\subsection{Conventional Approach}

Now that it becomes a common fact that the network resources/functionalities are on the verge of attaining their maximum performance, integrating them becomes unavoidable. Such integration will guarantee true pervasiveness. In future, rather than subscribing to individual services separately, subscribers will most likely subscribe to a service provided by integrating several services in different domains, which may include communications, caching, computing \cite{magurawalage2015aqua}, and control. The confluence of these resources will induce a remarkable transformation in the design philosophy of future networks \cite{wang2017survey}. Hence, combining 4C in the overall 5G, 6G, and beyond mobile networks design becomes necessary, especially for surmounting the challenges of emerging technologies.

To this end, the efforts in \cite{ndikumana2019joint} and \cite{kim20185g} focused on utilizing the network edges to achieve maximum network utility through the convergence of 4C. In particular, Ndikumana et al. \cite{ndikumana2019joint} proposed an integrated framework of 4C for managing big data in MEC. The framework enables big data computing and caching operations at the MEC servers, thus reducing end-to-end latency. These servers occupy the same cluster and actively collaborate to share the 4C resources. The technical rationale behind this is to: i) improve the backhaul network traffic, ii) maximize utilization of resources, and iii) minimize latency in the integrated 4C. The authors collaboratively optimized 4C to maximize bandwidth while minimizing latency under the constraints of the computing deadline, the local computational power of UEs, and MEC resources. Because of decision-making variables at multiple locations, the optimization problem, which is non-convex, was solved using a modified version of the block successive upper bound minimization (BSUM) approach. 

In contrast, Kim \cite{kim20185g} applied a hierarchical game-theoretic control algorithm for integrating resources. To be specific, the author devised a 5G network SBS, where a holistic control scheme characterizes the competitive and collaborative interactions among the communication, computing, and caching resources. This was realized by adopting game theory, which has to do with tactical interactions among several intelligent logical decision makers that systematically follow their objectives while maximizing the anticipated value of their payoffs. Due to its inherent ability to define the interactions among intelligent agents with conflicting objectives, game theory is applied to handle diverse competitive issues of network resources in wireless communications. Thus, the proposed approach considers: i) offloading decision at each UE, ii) radio splitting decision for data and content caching capacities, and iii) bandwidth allocation decision for each individual SBS. These decision issues require leveraging design principles, including self-interactivity, feasibility, and integral combination of various control algorithms, which depend on each other, to deal with conflicting performance benchmark under highly diverse 5G network circumstances.

In a design of 5G network SBS, where data caching, computation offloading processing, and mobile communication technologies are jointly utilized, a new control paradigm has to be employed to leverage the synergistic benefits of the integrated communication, computing, and caching operations in the SBSs. By doing so, different communication, computing, and caching characteristics can be captured to realize a promising solution under diverse 5G network circumstances. Therefore, as a control theory of several goal-oriented agents, game theory offers numerous promising solutions that optimize the overall performance of 5G networks; see \cite{kim20185g}.

In \cite{huo2016software}, Huo et al. relied on the principle of programmable control and caching, stemmed from SDN and ICN, respectively. Based on these premises, they proposed an integrated framework that systematically combines in-network caching, computation, and networking resources. These resources are centrally controlled and managed by utilizing SDN controller, thus monitoring both UEs and resources in the data plane. The framework enables dynamic orchestration of networking, caching, and computing resources to match the needs of future green networks. Conversely, \cite{chen2018joint} employed SDN to introduce an integrated framework of 4C. By fully considering the ability of the programmable control in SDN, Chen et al. proposed a framework that systematically converges networking, computation, and caching resources and allows the control functionality to dynamically orchestrate them. Unlike SDN, which programmably controls the switching devices’ forwarding function, this framework controls the data plane’s three-dimensional resources. It is also considered as service-oriented that supports general in-network services, thus different from content-oriented ICN with fixed in-network services. The study in \cite{zhang2018software} described such framework in Fig. 6, where the data plane is composed of caching, computing, and forwarding devices; the management/control modules for computing and caching resources are appropriately deployed at the control plane. The management plane bears the responsibility of monitoring and configuring the control functionality remotely by leveraging the SDN controller. The framework realizes an effective resource allocation and network orchestration by dynamically guiding various computing and content services to the corresponding service requesting UEs. 

\begin{figure}[t!]
	
	\includegraphics[width=90mm]{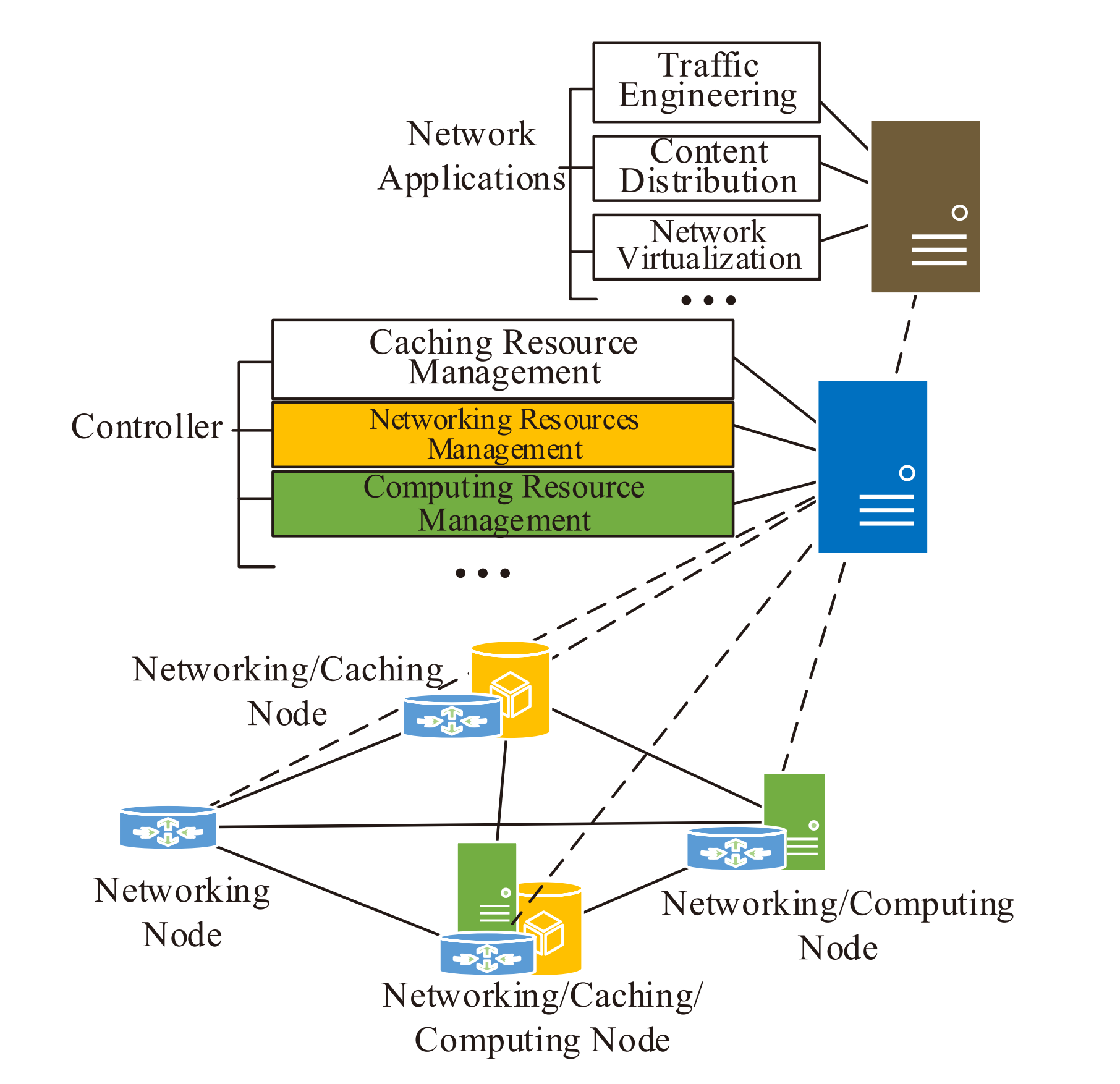}
	\caption{An integrated 4C framework \cite{zhang2018software}.}
	\label{fig5}
\end{figure}

\subsection{Intelligence for Resources Integration}

With the recent emergence of edge intelligence, intelligent control will inevitably interact with computing, communication, and caching functionalities at the network edge to promote the overall network performance and maximize end-to-end users' experience. Currently, there are two key factors in existing literature that motivate integrating/optimizing the edge resources with intelligent control, i.e., 1) complexity of mobile wireless networks and 2) exponential growth of data, as explored in the sequel.

\subsubsection{Complexity of Mobile Wireless Networks} Mobile wireless networks are becoming more complex nowadays. Such complex networks involve networking, computing, wireless communications, caching, and so on \cite{wang2020convergence}. Moreover, the incoming 6G networks will most likely be highly complex and dynamic. Conventional optimization algorithms with weighty mathematical models, such as gradient methods and Lagrangian duality, may not be desirable candidates for 6G. Hence, in AI-powered 6G networks, the parameters and frameworks will be optimized by utilizing the AI techniques rather than conventional tedious computation. Such techniques have good prospects of training and self-learning models for realizing network optimization in 6G networks, enabling network operators and providers to optimize the resources and parameters for enhanced QoS \cite{yang2020artificial}.

Wang et al. \cite{wang2019edge} raised some points spurring the convergence of intelligence with the communication, computing, and caching functionalities. Specifically, the effort pointed out that several existing efforts handled the resource allocation issues. Such efforts, in their assuming settings (mostly built upon game theory, convex optimization, etc.), realized quite impressive results. However, with specific use cases in MEC, such optimization approaches can be limited by: i) \textit{Uncertain Inputs}: here the assumption is that some relevant information factors are considered as inputs, unfortunately obtaining some of them remains a herculean task owing to randomness in wireless channels and privacy policies; ii) \textit{Dynamic Conditions}: the confluence of communication and computing resources dynamics are still under study; and iii) \textit{Temporal Isolation}: apart from Lyapunov optimization, most of those optimization problems overlook the lasting effect of current decisions on the allocation of resources, i.e., majority of traditional optimization algorithms are near-optimal or optimal only for a snapshot of the system in an extremely time-varying dynamic MEC framework. In short, diversified network devices requirements, wireless channels variations, different QoS needs, and much more complicate conventional optimization of 4C in 5G and 6G networks \cite{pham2020survey}. 

Besides, edge computing requires powerful optimization tools to address resource allocation challenges at various layers, including bandwidth, radio frequency, access jurisdiction, cache capacity, and CPU cycle frequency. AI solutions will surely handle such tasks \cite{deng2020edge}. Therefore, in 6G and beyond networks, a finest solution to the resource allocation optimization is attributed to AI (e.g., ML, DL, and FL techniques). Deng et al. \cite{deng2020edge} highlighted how to utilize the AI techniques at the edge to obtain more optimum solutions.

\subsubsection{Exponential Data Growth}

Today, we witness the unprecedented growth of data, created by different prevalent devices from mobile phones to industrial robots \cite{peltonen20206g}. In fact, billions of connected IoT and mobile devices generate massive volumes of data in zillions of bytes at the network edges. Motivated by this trend, the AI techniques have to be pushed to the network edges in order to completely unleash the capability of the edge big data \cite{zhou2019edge}. Hence, by leveraging an MEC platform and utilizing voluminous data distributed over myriad connected devices, the limitations of computing capability and finite data can be overcome at each device. The two critical and combined aspects in such systems are i) communicating between connected devices and edge servers and ii) learning from distributed data \cite{zhu2020toward}. Learning involves harnessing or altering existing knowledge and experience with the goal of improving a device or service center. Learning will bring numerous gains to 6G networks (including resource management, optimum network slicing, and so on), as per the applications requirements \cite{yang2020artificial}.

At the network edges, AI can exploit its incredible learning and reasoning capability to obtain significant information from data and perform decision-making, thus realizing intelligent management and maintenance of a network involving communication, computing, and caching resources \cite{wang2020convergence}. In a word, big data analytics and real-time processing require the distributed integration/optimization of the network edge resources and intelligent control at close proximity of the data sources.

\subsection{AI-based Approach}%mu2019latency,hamza2013survey,

 The thrust of recent studies in \cite{wang2019edge,he2018integrated} and \cite{li2018deep,hu2019ready,he2017integrated,hu2018mobility} focused on overcoming several identified issues of the i4C by harnessing the AI frontiers at the network edges. Wang et al \cite{wang2019edge} examined resource allocation optimization constraints in time-varying MEC systems and accordingly incorporated the DRL techniques (explicitly, Deep Q-Learning) and distributed DRL approaches into mobile edge systems. By coupling the DRL techniques and FL framework with MEC, the authors designed a framework to optimize communication, computing, and caching resources with intelligence at the edges. The 4C framework, termed In-Edge AI, intelligently harnesses the collaboration between the edge nodes and UEs to ensure better training and inference of the models through the exchange of learning parameters. Thus, performing dynamic network-level optimization and application-level enhancements meanwhile minimizing unnecessary communication burden. The gains are i) with DRL at the edges, information collecting, cognitive computing, and requests handling can be enabled, leading to an effective management of joint resources and ii) with FL framework, the DRL agents can be trained in a distributed way while: dramatically minimizing the quantity of data that should be uploaded over the wireless uplink channel, protecting the individual data security, responding cognitively to wireless networks' conditions and mobile communications environment, and fitting well with multiple mobile UEs in an effective mobile network.

 He et al. \cite{he2018integrated} focused on studying the dynamic nature of communication, computing, and caching resources and how to optimally allocate them using automatic decision-making intelligent control. The central idea is enabling efficient resource sharing for mobile social networks under the integrated resources scheme. In particular, the authors explored the sharp increase in trust-based social networks with the recent growth of D2D communications, MEC, and caching and proposed an integrated resources scheme. The integrated scheme faces resource allocation complications for UEs, especially in time-varying network resources conditions. Thus, when the dynamic trust values, wireless channel conditions, computational capabilities, and the cache status are jointly considered, the integrated network becomes more complicated, hence solving the problem by conventional optimization approach will be tedious. In such case, applying an intelligent control, based on the DRL approach, to automatically make resource allocation decisions by considering the networks conditions is a desirable solution. To that end, a Q-learning-based resource allocation strategy was applied to solve the optimized 4C problem without any explicit assumptions or simplification. In \cite{li2018deep}, Li et al. considered decent gains of RL in time-varying dynamic systems (such as multi-user MEC) and accordingly developed an RL-based optimization framework for handling the resource allocation in MEC. Specifically, the computation offloading decisions and resource allocation were jointly formulated as an optimization problem, which was solved by utilizing Q-learning and Deep Q Network (DQN).

%%%%%%%%%%%%%%%%%%%%%%%%%%%%%%%%%%%%%%%%%%%%%%%%%%%%%%%%%%%%%%%%%%%%%%%%%%%%%

\begin{table*}[t]
	\caption{Comparison of existing works on the i4C resources}
	\label{table2}
      \begin{center} 
		
		{\tiny\renewcommand{\arraystretch}{1}
			\resizebox{!}{.18\paperheight}{%
			
			\begin{tabular}{ | p{1.2cm} | p{1.1cm} | p{3.3cm}  |p{2cm} |p{1cm} |p{0.6cm} |}
				\hline

				\hline
				
				\hline \hline

				\textbf {Themes}	& \textbf{Networks} & \textbf{Key Contributions } &  \ \textbf{Objectives}&\textbf{Evaluation Method} &\textbf{Related Papers}\\

				\hline 
				
				\hline
				
				\hline  
				%%%%%%%%%%%%%%%%%%%%%%%%%%%%%%%%%%%%%%%%%%%%%%%%%%%%%%%%%%%%%%%%%%%%%%%%%%%%%%%%%%%%%%%%%%%%%
				
				&  Wireless Het-Nets
				& Proposing an integrated mechanism of 4C for processing big data in MEC.  &To increase bandwidth saving and lower latency. &Simulation. & \cite{ndikumana2019joint}\\

				\cline{2-6} 
				
				%%%%%%%%%%%%%%%%%%%%%%%%%%%%%%%%%%%%	

				&  5G mobile network
				& Introducing an integrated control scheme to harness synergies among the mobile communication, computing, and caching capabilities. &To maximize the 5G network performance. &Numerical and Simulation. & \cite{kim20185g}\\

				\cline{2-6} 
				
				%%%%%%%%%%%%%%%%%%%%%%%%%%%%%%%%%%%%

				&  Software defined networks
				& Proposing an integrated software defined networking, caching, and computing (SD-NCC) framework.  &To promote the system performance and satisfy the needs of diverse applications. &Simulation. & \cite{chen2018joint}\\

				\cline{2-6} 
				
				%%%%%%%%%%%%%%%%%%%%%%%%%%%%%%%%%%%%
				
				& Green wireless networks
				& Integrated framework for dynamic orchestration of network resources. &To satisfy the QoS requirements of various applications. &Simulation. & \cite{huo2016software}\\

				\cline{2-6} 
				
				%%%%%%%%%%%%%%%%%%%%%%%%%%%%%%%%%%%%
				
				&  Mobile UAV networks & Introducing an AI-based decision-making architecture for collaborative optimization of 4C and UAV team. &To maximize the network performance by utilizing the available resources. & Experiments. & \cite{hu2019ready}\\ 
				
				\cline{2-6}  
				
				%%%%%%%%%%%%%%%%%%%%%%%%%%%%%%%%%%%%

				i4C	&  Mobile networks
				& Integrating the DRL techniques and FL framework with mobile edge systems to optimize communication, computing, and caching resources with intelligence at the edge systems. &To promote content delivery and increase mobile QoS. &Experiment and Simulation. & \cite{wang2019edge}\\ 
				
				\cline{2-6} 
				
				%%%%%%%%%%%%%%%%%%%%%%%%%%%%%%%%%%%%

				&  Mobile wireless networks
				& Introducing a trust-based social networks framework with D2D communication, in-network caching, and MEC. &To maximize the efficiency and security of mobile social networks. &Simulation. & \cite{he2018integrated}\\ 
				
				\cline{2-6} 
				
				%%%%%%%%%%%%%%%%%%%%%%%%%%%%%%%%%%%%
				
				&  Wireless Het-Nets
				& Integrated system for multi-user computation offloading and resource allocation. &To minimize the total delay and energy consumption. &Simulation. & \cite{li2018deep}\\ 
				
				\cline{2-6} 
				
				%%%%%%%%%%%%%%%%%%%%%%%%%%%%%%%%%%%%
				
				&  Vehicular networks
				& Integrated framework for dynamic orchestration of networking, computing, and caching resources. &To increase the performance of future generation vehicular networks. &Simulation. & \cite{he2017integrated}\\ 
				
				\cline{2-6} 
				
				%%%%%%%%%%%%%%%%%%%%%%%%%%%%%%%%%%%%
				
				&  Vehicular networks
				& Developing a joint optimal resource allocation framework for vehicular networks. &To realize operational excellence and increase the cost efficiency in vehicular networks. &Numerical. & \cite{hu2018mobility}\\ 
				
				\cline{2-6} 
				
				%%%%%%%%%%%%%%%%%%%%%%%%%%%%%%%%%%%%

				\hline 
				
				\hline
				
				\hline  \hline
				
			\end{tabular}

   }}
   \end{center}
\end{table*}

%%%%%%%%%%%%%%%%%%%%%%%%%%%%%%%%%%%%%%%%%%%%%%%%%%%%%%%%%%%%%%%%%%%%%%%%%%%%%

%\begin{figure}[t!]
%	
%	\includegraphics[width=90mm]{F3.jpg}
%	\caption{Three-node MEC framework with joint cooperation of communication and computation resources. The broken and solid lines show tasks offloading to the helper (for computing cooperation) and to the AP (via the helper's communication cooperation as a relay), respectively \cite{hao2018energy}, [181].}
%	\label{fig3}
%\end{figure}

%\begin{figure}[t!]
%	
%	\includegraphics[width=90mm]{F4.jpg}
%	\caption{Four-slots protocols with joint cooperation of communication and computation resources in MEC \cite{hao2018energy}, [181].}
%	\label{fig4}
%\end{figure}

%%%%%%%%%%%%%%%%%%%%%%%%%%%%%%%%%%%%%%%%%%%%%%%%%%%%%%%%%%%%%%%%%%%%%%%%%%%%%%%%%%%%%%%%%%%%%%%%%%%%%

\begin{table*}[t]
	
	\centering
	\caption{Summary of existing efforts on the integration of communication and computing.}
	\label{t3}
	
	\begin{adjustbox}{width=1\textwidth}
		\begin{tabular}{||c|c|c|c|c|c|c|c|}
			\hline

			\hline
			
			\hline \hline
			\multirow{1}{*}{} &
			\multirow{1}{*}{} &
			\multirow{1}{*}{} &
			\multicolumn{3}{c}{\textbf{Objective} } &
			
			\multicolumn{1}{|c}{ } &
			\multicolumn{1}{|c|}{ } \\
			
			\cline{4-6}

			\textbf {Integration}	& \textbf{Network} & \textbf{Main } & Latency   & Energy   & Other Performance   &  \textbf{Evaluation  } & \textbf{Reference}  \\
			
			&  &  \textbf{Contribution }& Minimization &  Saving &  Metrics   &  \textbf{Method} & \\

			\hline 
			
			\hline
			
			\hline  
			%%%%%%%%%%%%%%%%%%%%%%%%%%%%%%%%%%%%%%%%%%%%%%%%%%%%%%%%%%%%%%%%%%%%%%%%%%%%%%%%%%%%%%%%%%%%%	

			& Wireless   & Formulating a joint     
			&  & &   & &\\
			
			%				\cline{2-8}
			
			& Het-Nets &communication and computing & $\surd$  &   &   & Numerical &\cite{barbarossa2014communicating}\\

			&  &  resources optimization problem  &  &  &  &  &\\

			\cline{2-8} 
			
			%%%%%%%%%%%%%%%%%%%%%%%%%%%%%%%%%%%%%%%%%%%%%%%%%%%%%%%%%%%%%%%%%%%%%%%%%%%%%%%%%%%%%%%%%%%%%
			
			& Mobile     & Proposing a joint communication    &  & &   & &\\
			
			%				\cline{2-8}
			
			& wireless & and computation resource    & $\surd$  &   &   & Numerical &\cite{ren2018latency}\\

			& network & allocation for a TDMA-based   &  &  &  &  &\\
			
			&  &   multi-user MECO system  &  &  &  &  &\\

			\cline{2-8}
			
			%%%%%%%%%%%%%%%%%%%%%%%%%%%%%%%%%%%%%%%%%%%%%%%%%%%%%%%%%%%%%%%%%%%%%%%%%%%%%%%%%%%%%%%%%%%%%
			
			& Mobile     & Proposing a joint   &  & &   & &\\
			
			%				\cline{2-8}
			
			& wireless & communication and computation    & $\surd$  & $\surd$  &   & Simulation &\cite{li2019communication}\\
			
			&   network   &   cooperation approach  &  & &   & &\\

			\cline{2-8}
			%%%%%%%%%%%%%%%%%%%%%%%%%%%%%%%%%%%%%%%%%%%%
			
			& Mobile     & Proposing a joint   &  & &   &Numerical &\\
			
			%				\cline{2-8}
			
			& wireless & communication and computation   &   &  $\surd$ &  &  and  &\cite{cao2018joint}\\
			
			&   network   & cooperation approach  &  & &   & simulation &\\

			\cline{2-8}

			%%%%%%%%%%%%%%%%%%%%%%%%%%%%%%%%%%%%%%%%%%%%%%%%%%%%%%%%%%%%%%%%%%%%%%%%%%%%%%%%%%%%%%%%%%%%%
			
			Communication	& Wireless   & Presenting an integrated     
			&  & &   &Numerical   &\\
			
			%				\cline{2-8}
			
			and & Het-Nets &framework for computation offloading &   & $\surd$  &   & and &\cite{zhang2017energy}\\
			
			Computing &  &and resource allocation in MEC &   &   &  & simulation &\\

			\cline{2-8} 
			
			%%%%%%%%%%%%%%%%%%%%%%%%%%%%%%%%%%%%%%%%%%%%%%%%%%%%%%%%%%%%%%%%%%%%%%%%%%%%%%%%%%%%%%%%%%%%%
			
			& Mobile     & Hybrid pre-coding with    &  & &   & &\\
			
			%				\cline{2-8}
			
			& wireless &  communication and computational    &   &  $\surd$ &   & Simulation &\cite{ge2018joint}\\
			
			&   network   &  capabilities algorithm &  & &   & &\\

			\cline{2-8}

			%%%%%%%%%%%%%%%%%%%%%%%%%%%%%%%%%%%%%%%%%%%%%%%%%%%%

			& Wireless   & Formulating a joint      &  & &   & &\\
			
			%				\cline{2-8}
			
			& Het-Nets & optimization for transmission &   &  &  $\surd$ & Simulation &\cite{chen2017joint}\\

			&  & and processing delays &  &  &  &  &\\

			\cline{2-8}

			%%%%%%%%%%%%%%%%%%%%%%%%%%%%%%%%%%%%%%%%%%%%%%%%%%%%%%%%%%%%%%%%%%%%%%%%%%%%%%%%%%%%%%%%%%%%%
			
			& Wireless   & Integrated framework of        &  & &   & &\\
			
			%				\cline{2-8}
			
			& Het-Nets & VFC for communication &   &  &  $\surd$ & Simulation &\cite{hou2016vehicular}\\

			&  & and computing resources &  &  &  &  &\\

			\cline{2-8} 
			
			%%%%%%%%%%%%%%%%%%%%%%%%%%%%%%%%%%%%%%%%%%%%%%%%%%%%%%%%%%%%%%%%%%%%%%%%%%%%%%%%%%%%%%%%%%%%%
			
			& Wireless   & Proposing an MEC collaborative  
			&  & &   & &\\
			
			%				\cline{2-8}
			
			& Het-Nets & architecture for resource &   &  & $\surd$   & Simulation &\cite{yang2017computation}\\

			&  & sharing among MEC-BSs in UDN  &  &  &  &  &\\

			\cline{2-8}

			\hline 
			
			\hline
			
			\hline  \hline

		\end{tabular}

	\end{adjustbox}

\end{table*}

%%%%%%%%%%%%%%%%%%%%%%%%%%%%%%%%%%%%%%%%%%%%%%%%%%%%%%%%%%%%%%%%%%%%%%%%%%%%%%%%%%%%%%%%%%%%%%%%%%%%%%%%%%%%%%%%%

%%%%%%%%%%%%%%%%%%%%%%%%%%%%%%%%%%%%%%%%%%%%%%%%%%%%%%%%%%%%%%%%%%%%%%%%%%%%%%%%%%%%%%%%%%

\begin{table*}[htbp]
	
	\centering
	\caption{Summary of existing efforts on the integration of communication and control.}
	\label{t3}
	
	\begin{adjustbox}{width=1\textwidth}
		\begin{tabular}{||c|c|c|c|c|c|c|}
			\hline

			\hline
			
			\hline \hline
			\multirow{1}{*}{} &
			\multirow{1}{*}{} &
			\multirow{1}{*}{} &
			\multicolumn{2}{c}{\textbf{Objective} } &
			
			\multicolumn{1}{|c}{ } &
			\multicolumn{1}{|c|}{ } \\
			
			\cline{4-5}

			\textbf {Integration}	& \textbf{Network} & \textbf{Main } & Latency  & Other Performance   &   \textbf{Evaluation  } & \textbf{Reference}  \\
			
			&  &  \textbf{Contribution }& Minimization &    Metrics   &  \textbf{Method} & \\

			\hline 
			
			\hline
			
			\hline  
			%%%%%%%%%%%%%%%%%%%%%%%%%%%%%%%%%%%%%%%%%%%%%%%%%%%%%%%%%%%%%%%%%%%%%%%%%%%%%%%%%%%%%%%%%%%%%
			
			& V2V      & Proposing an integrated control        &    & &   &\\

			& wireless     & system and V2V wireless & $\surd$ &  &Simulation &\cite{zeng2019joint}\\

			& networks & communication co-design framework  &   &  &  &\\
			
			\cline{2-7}

			%%%%%%%%%%%%%%%%%%%%%%%%%%%%%%%%%%%%%%%%%%%%%%%%%%%%%%%%%%%%%%%%%%%%%%%%%%%%%%%%%%%%%%%%%%%%%
			
			& V2V      & Proposing an integrated control        &    & &   &\\

			& wireless     & system and V2V wireless & $\surd$ &  &Simulation &\cite{zeng2018integrated}\\

			& networks & communication co-design framework  &   &  &  &\\
			
			\cline{2-7}

			%%%%%%%%%%%%%%%%%%%%%%%%%%%%%%%%%%%%%%%%%%%%%%%%%%%%%%%%%%%%%%%%%%%%%%%%%%%%%%%%%%%%%%%%%%%%%

			& Wireless       & Applying a Smith        &    & &   &\\

			& vehicular      & predictor  & $\surd$ &  &Experimental  &\cite{xing2018smith}\\

			& networks &   &   &  &  &\\
			
			\cline{2-7}

			%%%%%%%%%%%%%%%%%%%%%%%%%%%%%%%%%%%%%%%%%%%%%%%%%%%%%%%%%%%%%%%%%%%%%%%%%%%%%%%%%%%%%%%%%%%%%

			& Wireless       & Joint control and        &    & &   &\\

			&   networks    & communication system design  &$\surd$ &  &Simulation  &\cite{zeng2018wireless}\\

			\cline{2-7}

			%%%%%%%%%%%%%%%%%%%%%%%%%%%%%%%%%%%%%%%%%%%

			& Wireless   & Proposing a joint  control    &   &  &   &\\

			& NCS &   and communication design &   & $\surd$ & Simulation &\cite{chamaken2010joint}\\

			\cline{2-7}
			
			%%%%%%%%%%%%%%%%%%%%%%%%%%%%%%%%%%%%%%%%%%%%%%%%%%%%%%%%%%%%%%%%%%%%%%%%%%%%%%%%%%%%%%%%%%%%%
			
			Communication	& Wireless    & Proposing three self- & &  &   &\\

			and	& NCS & triggered control strategies &   &  $\surd$ & Numerical &\cite{akashi2018self}\\

			\cline{2-7}

			%%%%%%%%%%%%%%%%%%%%%%%%%%%%%%%%%%%%%%%%%%%%%%%%%%%%%%%%%%%%%%%%%%%%%%%%%%%%%%%%%%%%%%%%%%%%%
			Control	& Wireless    & Joint UAV trajectory and  &   &  & &\\

			& network  & power control scheme &   & $\surd$ &Numerical &\cite{huang2018cognitive}\\

			\cline{2-7}

			%%%%%%%%%%%%%%%%%%%%%%%%%%%%%%%%%%%%%%%%%%%%%%%%%%%%%%%%%%%%%%%%%%%%%%%%%%%%%%%%%%%%%%%%%%%%%

			& Wireless    & Proposing an adaptive      &    & &   &\\

			& cellular   & distributed power &  & $\surd$ &Simulation &\cite{payasi2015negotiation}\\

			& networks & control algorithm  &   &  &  &\\
			
			\cline{2-7}

			%%%%%%%%%%%%%%%%%%%%%%%%%%%%%%%%%%%%%%%%%%%%%%%%%%%%%%%%%%%%%%%%%%%%%%%%%%%%%%%%%%%%%%%%%%%%%
			
			& Intervehicle     & Proposing a novel       &    & &   &\\

			& communication    & adaptive switched &  & $\surd$ &Simulation &\cite{abou2017adaptive}\\

			& networks & control algorithm  &   &  &  &\\

			\cline{2-7}

			\hline 
			
			\hline
			
			\hline  \hline

		\end{tabular}

	\end{adjustbox}

\end{table*}

%%%%%%%%%%%%%%%%%%%%%%%%%%%%%%%%%%%%%%%%%%%%%%%%%%%%%%%%%%%%%%%%%%%%%%%%%%%%%%%%%%%%%%%%%%%%%%%%%%%%%%%%%%%%%%%%%

On the other hand, the need to process complex tasks, e.g., airborne/aerial imagery, precision target identification, and adaptive cruise, in the air garners much attention today. Unfortunately, multiple factors complicate the real-time interactions between an MEC system and a UAV. For example, UAV has a short battery lifespan, which may drain fast while interacting with MEC; it may also suffer from insufficient computing and storage resources due to its miniaturized body.  Considering these factors, Hu et al. \cite{hu2019ready} proposed a framework to collaboratively optimize the communication, computing, and caching resources and the UAV swamp with AI-based decision-making scheme. To efficiently process complex tasks in the air by enabling flexible interaction between MEC and UVA, the framework considers UAV swamp deployments both in real-time based on real-time perception and in advance based on historical data mining. The swamp forms a dynamic resource pool based on UAV collaborations in multi-task-offloading scenarios. At the same time, their resources can be dynamically and flexibly allocated to one another so as to balance the resources utilization.

The efforts in \cite{he2017integrated} and \cite{hu2018mobility} focused on addressing wide-ranging issues of vehicular networks by optimizing 4C functionalities. In \cite{he2017integrated}, He et al. optimized 4C to fulfill the requirements of emerging vehicular networks. Specifically, they introduced an integrated system to dynamically orchestrate networking, computing, and caching functionalities. This was realized based on the principle of programmable control and the concept of information centricity, derived from SDN and ICN, respectively. Then, resource allocation strategy was constructed as an optimization problem while considering the respective gains of computing, caching, and networking functionalities. Due to the high complexity of the system, the authors observed striking features of RL and offered a novel DRL technique for the optimization problem. In \cite{hu2018mobility}, the hard service deadline constraints and the vehicle's mobility were considered to design the resource allocation policy. To be more specific, the authors jointly optimized communication, computing, and caching design problem to maximize the cost efficiency and realize operational excellence of vehicular networks. The formulated problem was solved by developing deep Q-learning based algorithm with multi-timescale framework. Furthermore, the mobility-aware reward estimation for the large timescale model was proposed to reduce the complexity caused by the large action space. Table 2 compares the contributions of existing efforts on the i4C.

In summary, AI shows great potential in a mobile network environment, where different network challenges can be dealt with by jointly optimizing communication, computing, caching with intelligent control (e.g., DRL). Indeed, the DRL approach achieves breakthroughs due to its ability to make optimal resource allocation decisions, especially in a highly time-varying dynamic systems. The performance of the DRL approach in the 4C optimization can be demonstrated by using TensorFlow in computer simulations. TensorFlow is an ML system operating in large-scale and effectively applied to wireless Het-Nets. Specifically, TensorFlow utilizes unified dataflow graphs to address the computations in algorithms, states, and actions. It can map the dataflow graph nodes across several machines in a group and also in a machine across various computing devices, such as general-purpose GPUs, multi-core CPUs, and custom designed ASICs (aka Tensor Processing Units). Being an open source, TensorFlow can provide developers with flexibility and enable them to conduct experiments with innovative optimizations and training algorithms according to prior parameter server designs. Therefore, it supports diverse applications needing training and inference algorithms on advanced deep neural networks; see \cite{hu2018mobility,abadi2016tensorflowb,abadi2016tensorflow}.
%%%%%%%%%%%%%%%%%%%%%%%%%%%%%%%%%%%%%%%%%%%%%%%%%%%%%%%%%%%%%%%%%%%%%%%%%%%%%%%%%%%%%%%%%%%%%%%%%%%%%

\begin{table*}[htbp]
	
	\centering
	\caption{Summary of existing efforts on the integration of computing and caching.}
	\label{t3}
	
	\begin{adjustbox}{width=1\textwidth}
		\begin{tabular}{||c|c|c|c|c|c|c|c|}
			\hline

			\hline
			
			\hline \hline
			\multirow{1}{*}{} &
			\multirow{1}{*}{} &
			\multirow{1}{*}{} &
			\multicolumn{3}{c}{\textbf{Objective} } &
			
			\multicolumn{1}{|c}{ } &
			\multicolumn{1}{|c|}{ } \\
			
			\cline{4-6}

			\textbf {Integration}	& \textbf{Network} & \textbf{Main } & Latency   & Energy   & Other Performance   &  \textbf{Evaluation  } & \textbf{Reference}  \\
			
			&  &  \textbf{Contribution }& Minimization &  Saving &  Metrics   &  \textbf{Method} & \\

			\hline 
			
			\hline
			
			\hline  
			
			%%%%%%%%%%%%%%%%%%%%%%%%%%%%%%%%%%%%%%%%%%%%%%%%%%%%%%%%%%%%%%%%%%%%%%%%%%%%%%%%%%%%%%%%%%%%%
			
			& Wireless   & Proposing optimal  &  & &   & &\\
			
			%				\cline{2-8}
			
			& Het-Nets & offloading with caching & $\surd$  &  &   & Simulation &\cite{yu2018computation}\\

			&  & enhancement scheme (OOCS)  &  &  &  &  &\\

			\cline{2-8} 
			
			%%%%%%%%%%%%%%%%%%%%%%%%%%%%%%%%%%%%%%%%%%%%%%%%%%%%%%%%%%%%%%%%%%%%%%%%%%%%%%%%%%%%%%%%%%%%%

			& Wireless   & Proposing a data allocation   &  & &   & &\\
			
			%				\cline{2-8}
			
			& Het-Nets & algorithm and an offloading &  $\surd$ &  &   & Simulation &\cite{fan2016terminalbooster}\\

			&  & scheduling algorithm  &  &  &  &  &\\

			\cline{2-8} 
			
			%%%%%%%%%%%%%%%%%%%%%%%%%%%%%%%%%%%%%%%%%%%%%%%%%%%%%%%%%%%%%%%%%%%%%%%%%%%%%%%%%%%%%%%%%%%%%
			
			& Wireless   &Formulating an integrated model     &  & &   & &\\
			
			%				\cline{2-8}
			
			& Het-Nets & of computation offloading, caching, &  $\surd$ &  &   & Simulation &\cite{zhang2018joint}\\

			&  & and resource allocation  &  &  &  &  &\\

			\cline{2-8} 
			
			%%%%%%%%%%%%%%%%%%%%%%%%%%%%%%%%%%%%%%%%%%%%%%%%%%%%%%%%%%%%%%%%%%%%%%%%%%%%%%%%%%%%%%%%%%%%%
			
			& Wireless   & Task caching and      &  & &   & &\\
			
			%				\cline{2-8}
			
			& Het-Nets & computation offloading scheme & $\surd$  & $\surd$&    & Simulation &\cite{hao2018energy}\\

			\cline{2-8} 
			
			%%%%%%%%%%%%%%%%%%%%%%%%%%%%%%%%%%%%%%%%%%%%%%%%%%%%%%%%%%%%%%%%%%%%%%%%%%%%%%%%%%%%%%%%%%%%%
			
			Computing	& Wireless   & Proposing OREO to jointly      &  & &   & &\\
			
			%				\cline{2-8}
			
			and	& Het-Nets & optimize dynamic service &  $\surd$ &$\surd$ &    & Simulation &\cite{xu2018joint}\\

			Caching	&  & caching and computation offloading   &  &  &  &  &\\

			\cline{2-8} 
			
			%%%%%%%%%%%%%%%%%%%%%%%%%%%%%%%%%%%%%%%%%%%%%%%%%%%%%%%%%%%%%%%%%%%%%%%%%%%%%%%%%%%%%%%%%%%%%
			
			& Wireless   & Constructing a joint       &  & &   & &\\
			
			%				\cline{2-8}
			
			& Het-Nets & computation offloading scheduling & $\surd$  &$\surd$ &    & Simulation &\cite{liu2018computation}\\

			&  &  and caching scheme  &  &  &  &  &\\

			\cline{2-8} 
			
			%%%%%%%%%%%%%%%%%%%%%%%%%%%%%%%%%%%%%%%%%%%%%%%%%%%%%%%%%%%%%%%%%%%%%%%%%%%%%%%%%%%%%%%%%%%%%

			& Wireless   & Computation and caching      &  & &   & &\\
			
			%				\cline{2-8}
			
			& Het-Nets & resources allocation in MEC & &  &  $\surd$   & Simulation &\cite{ndikumana2017collaborative}\\

			\cline{2-8}

			%%%%%%%%%%%%%%%%%%%%%%%%%%%%%%%%%%%%%%%%%%%%%%%%%%%%%%%%%%%%%%%%%%%%%%%%%%%%%%%%%%%%%%%%%%%%%
			
			& Wireless   & Formulating a joint collaborative        &  & &   & &\\
			
			%				\cline{2-8}
			
			& Het-Nets & caching and processing  &   &  &  $\surd$ & Simulation &\cite{tran2017collaborative}\\

			&  & problem as an ILP &  &  &  &  &\\

			\cline{2-8} 
			
			%%%%%%%%%%%%%%%%%%%%%%%%%%%%%%%%%%%%%%%%%%%%%%%%%%%%%%%%%%%%%%%%%%%%%%%%%%%%%%%%%%%%%%%%%%%%%
			
			& Wireless   & Proposing a joint  
			&  & &   & &\\
			
			%				\cline{2-8}
			
			& Het-Nets &  collaborative caching and   &   &  & $\surd$  & Simulation &\cite{tran2018adaptive}\\

			&  & processing framework  &  &  &  &  &\\

			\cline{2-8} 
			
			%%%%%%%%%%%%%%%%%%%%%%%%%%%%%%%%%%%%%%%%%%%%%%%%%%%%%%%%%%%%%%%%%%%%%%%%%%%%%%%%%%%%%%%%%%%%%

			\hline 
			
			\hline
			
			\hline  \hline

		\end{tabular}

	\end{adjustbox}

\end{table*}

%%%%%%%%%%%%%%%%%%%%%%%%%%%%%%%%%%%%%%%%%%%%%%%%%%%%%%%%%%%%%%%%%%%%%%%%%%%%%%%%%%%%%%%%%%%%%%%%%%%%%%%%%%%%%%%%%

%%%%%%%%%%%%%%%%%%%%%%%%%%%%%%%%%%%%%%%%%%%%%%%%%%%%%%%%%%%%%%%%%%%%%%%%%%%%%%%%%%%%%%%%%%%%%%%%%%%%%

\begin{table*}[t]
	
	\centering
	\caption{Summary of existing efforts on the integration of communication, computing, and caching.}
	\label{table2}
	\begin{center}
		{\tiny\renewcommand{\arraystretch}{1}
			\resizebox{!}{.2\paperheight}{%

		\begin{tabular}{||c|c|c|c|c|c|c|c|}
			\hline

			\hline
			
			\hline \hline
			\multirow{1}{*}{} &
			\multirow{1}{*}{} &
			\multirow{1}{*}{} &
			\multicolumn{3}{c}{\textbf{Objective} } &
			
			\multicolumn{1}{|c}{ } &
			\multicolumn{1}{|c|}{ } \\
			
			\cline{4-6}

			\textbf {Integration}	& \textbf{Network} & \textbf{Main } & Latency  & Energy   & Other	Performance     &  \textbf{Evaluation  } & \textbf{Reference}  \\
			
			&  &  \textbf{Contribution }& Minimization &  Saving & Metrics   &  \textbf{Method} & \\

			\hline 
			
			\hline
			
			\hline  
			%%%%%%%%%%%%%%%%%%%%%%%%%%%%%%%%%%%%%%%%%%%%%%%%%%%%%%%%%%%%%%%%%%%%%%%%%%%%%%%%%%%%%%%%%%%%%

			%%%%%%%%%%%%%%%%%%%%%%%%%%%%%%%%%%%%	
			
			& Edge  & Proposing a joint optimization &  &  &   & &\\
			
			%				\cline{2-8}
			
			&  -cloud  &   of communication, computing,  & $\surd$ &  &   & Simulation &\cite{chen2018edge}\\

			&    &   and caching on edge cloud, & &  &   &  &\\
			
			&    &   called Edge-CoCaCo & &  &   &  &\\

			\cline{2-8}
			
			%%%%%%%%%%%%%%%%%%%%%%%%%%%%%%%%%%%%%

			& Mobile  & Devising a joint   &  &  &  & &\\
			
			%				\cline{2-8}
			
			&  vehicular  &   communication, computing, and  &$\surd$ &  &   &Simulation  &\cite{kazmi2019infotainment}\\

			&   Networks & caching system model & &  &   &  &\\

			\cline{2-8}

			%%%%%%%%%%%%%%%%%%%%%%%%%%%%%%%%%%%%%%%%%%%%%%%%%%%%%%%%%%%%%%%%%%%%%%%%%%%%%%%%%%%%%%%%%%%%%
			
			& Virtualized  & Proposing an Air-ground  &  &  &   &Numerical  &\\
			
			%				\cline{2-8}
			
			&  Network  &   integrated MEC architecture & $\surd$ &  &   & and simulation & \cite{zhou2018air}\\

			\cline{2-8} 
			%%%%%%%%%%%%%%%%%%%%%%%%%%%%%%%%%%%%%%%%%%%%%%%%%%%%%%%%%%%%%%%%%%%%%%%%%%%%%%%%%%%%%%%%%%%%%
			
			& Wireless   & Proposing Hybrid IoT to enable   &  &  &   &  &\\
			
			%				\cline{2-8}
			
			&  Het-Nets  &   efficient transmission, computing, & $\surd$ & $\surd$ &   &  Numerical&\cite{qian2019hybridiot}\\
			
			&    &   and caching of big data & &  &   & and simulation &\\

			\cline{2-8}

			%%%%%%%%%%%%%%%%%%%%%%%%%%%%%%%%%%%%%%%%%%%%%%%%%%%%%%%%%%%%%%%%%%%%%%%%%%%%%%%%%%%%%%%%%%%%%
			
			& Mobile     & Developing implementation     &  &  &   &  &\\
			
			%				\cline{2-8}
			
			&  wireless  &  framework for mobile  &$\surd$ & $\surd$ &   & Numerical  &\cite{sun2018communication}\\
			
			&   network &   VR delivery & &  &   &  &\\

			\cline{2-8} 
			%%%%%%%%%%%%%%%%%%%%%%%%%%%%%%%%%%%%%%%%%%%%%%%%%%%%%%%%%%%%%%%%%%%%%%%%%%%%%%%%%%%%%%%%%%%%%
			
			& Mobile     & Presenting a novel   &  &  &   &  &\\
			
			%				\cline{2-8}
			
			&  wireless  &  MEC-based mobile VR  &$\surd$& $\surd$ &   & Numerical &\cite{sun2019communications}\\
			
			&   network &   delivery framework & &  &   &  &\\

			\cline{2-8} 
			
			%%%%%%%%%%%%%%%%%%%%%%%%%%%%%%%%%%%%%%%%%%%%%%%%%%%%%%%%%%%%%%%%%%%%%%%%%%%%%%%%%%%%%%%%%%%%%
			
			& F-RANs     & Enabling a joint caching and    &  &  &   &  &\\
			
			%				\cline{2-8}
			
			&    &   computing policy using the  & $\surd$& $\surd$ &   & Numerical &\cite{dang2019joint}\\
			
			&    &   communication, computing, and caching & &  &   &  &\\

			&    &   resource allocation problem & &  &   &  &\\
			
			\cline{2-8} 
			
			%%%%%%%%%%%%%%%%%%%%%%%%%%%%%%%%%%%%%%%

			%%%%%%%%%%%%%%%%%%%%%%%%%%%%%%%%%%%%%%%%%%%%%%%%%%%%%%%%%%%%%%%%%%%%%%%%%%%%%%%%%%%%%%%%%%%%%

			%%%%%%%%%%%%%%%%%%%%%%%%%%%%%%%%%%%%%%%%%%%%%%%%%%%%%%%%%%%%%%%%%%%%%%%%%%%%%%%%%%%%%%%%%%%%%
			
			Communication,	& Multiuser  & Formulating joint  &   & &   & Analytical/&\\
			
			%				\cline{2-8}
			
			Computing,& MEC-based & optimization of computing & &  &  $\surd$  & Numerical &\\

			and Caching	& wireless network & and caching policy &  &  & &  &\cite{sun2020bandwidth}\\

			\cline{2-8}    
			
			%%%%%%%%%%%%%%%%%%%%%%%%%%%%%%%%%%%%%%%%%%%%%%%%%%%%%%%%%%%%%%%%%%%%%%%%%%%%%%%%%%%%%%%%%%%%%

			%%%%%%%%%%%%%%%%%%%%%%%%%%%%%%%%%%%%%%%%%%%%%%%%%%%%%%%%%%%%%%%%%%%%%%%%%%%%%%%%%%%%%%%%%%%%%
			
			& Wireless   & Formulating a joint computation    &   & &   & &\\
			
			%				\cline{2-8}
			
			& Het-Nets  & offloading, spectrum resource and & &  &   &  &\\

			&  & computation resource allocation,&  &  & $\surd$ & Simulation &\cite{wang2017computation}\\

			&  & and content caching optimization  &  & 	 &  & & \\

			\cline{2-8}

			%%%%%%%%%%%%%%%%%%%%%%%%%%%%%%%%%%%%%%%%%%%%%%%%%%%%%%%%%%%%%%%%%%%%%%%%%%%%%%%%%%%%%%%%%%%%%
			
			& Wireless    & Formulating a joint computation      &  & &   & &\\
			
			%				\cline{2-8}
			
			& Het-Nets &  offloading, resource allocation, &   &  &  $\surd$ & Simulation &\cite{wang2017joint}\\

			&  & and content caching optimization&  &  &  &  &\\

			\cline{2-8}

			%%%%%%%%%%%%%%%%%%%%%%%%%%%%%%%%%%%%%%%%%%%%%%%%%%%%%%%%%%%%%%%%%%%%%%%%%%%%%%%%%%%%%%%%%%%%%
			
			& Virtualized    & Designing a novel information   &   & &   & &\\
			
			%				\cline{2-8}
			
			&  Het-Nets &  -centric Het-Nets framework &  &  & $\surd$  & Simulation & \cite{zhou2017resource}\\

			\cline{2-8} 
			
			%%%%%%%%%%%%%%%%%%%%%%%%%%%%%%%%%%%%%%%%%%%%%%%%%%%%%%%%%%%%%%%%%%%%%%%%%%%%%%%%%%%%%%%%%%%%%
			
			& Virtualized    & Designing a novel information    &  & &   & &\\
			
			%				\cline{2-8}
			
			&  Het-Nets &  -centric Het-Nets framework &  &  & $\surd$  & Simulation &\cite{zhou2017virtual}\\
			
			&   &  for sharing communication, & &  &   &  &\\
			
			&   &  computing, and caching resources & &  &   &  &\\

			\cline{2-8} 
			
			%%%%%%%%%%%%%%%%%%%%%%%%%%%%%%%%%%%%%%%%%%%%%%%%%%%%%%%%%%%%%%%%%%%%%%%%%%%%%%%%%%%%%%%%%%%%%

			\hline 
			
			\hline
			
			\hline  \hline

		\end{tabular}

	}}

\end{center}

\end{table*}

%%%%%%%%%%%%%%%%%%%%%%%%%%%%%%%%%%%%%%%%%%%%%%%%%%%%%%%%%%%%%%%%%%%%%%%%%%%%%%%%%%%%%%%%%%%%%%%%%%%%%

\subsection{Classification of the Integration Approaches}

Previously, several studies integrated key components/parts of 4C. The contributions of such efforts improved wireless networks performance and pave the way for the emergence and thriving of the i4C. This subsection dedicates itself to these essential concepts, providing good insights into the prospects of converging 4C in 5G, 6G, and beyond networks. Therefore, we focus on different approaches to i4C, classifying various works that converged or explored these four underlying functionalities.

Specifically, we classify the existing studies pertaining to i4C into integration of: 1) communication, computing, and caching; 2) communication, computing, and control; 3) communication and computing; 4) communication and caching; 5) communication and control; 6) computing and caching; 7) computing and control; and 8) caching and control. However, due to limited space, we summarize the contributions of some of these efforts based their objectives (i.e., time saving, energy saving, and other metrics) in Tables 3, 4, 5, and 6. Thus, this survey does not cover all existing literature related to i4C.

%\begin{figure}[t!]
%	
%	\includegraphics[width=80mm]{F5.jpg}
%	\caption{An architecture of caching framework with an MBS and multiple SBSs [201].}
%	\label{fig5}
%\end{figure}

In particular, the focus of discussion in Table 3 lies in the convergence of communication and computing, where the major thrust targeted low latency, energy minimization, and other key metrics. Table 4 presents the summary of existing studies on the integration of communication and control with focus on realizing low latency and other metrics. The themes of the efforts summarized in Table 5 lies in coupling computing and caching with the aim of realizing low latency, low energy consumption, latency and energy savings, and other performance metrics. Table 6 summarizes some existing efforts devoted to integrating communication, computing, and caching with the goals of attaining ultra-low latency, low energy consumption, low latency and energy savings, and other performance metrics.

\subsection{Integration of Sensing and Communication (ISAC)}
 In 6G wireless networks, various intensive computing services are expected to pop up. Distributed computing plays the role of a key 6G enabler that collectively utilizes pervasive sensing, communication, and computing functionalities in UEs, MEC/edge servers, and network nodes. With distributed computing, computation-intensive tasks can be partitioned into several subtasks and allocated to various network nodes for parallel collaborative computing. For instance, wireless distributed learning and reasoning could be employed to intelligently forecast future traffic pattern, network bottleneck, and resource availability based on archived and real-time big data in mobile networks\cite{zhou2020service}. Such data can be complex and enormous and have pivotal roles to play in combining sensing and communication. Data are exchanged among numerous UEs and also between the edge/cloud servers and UEs, including connected IoT devices, IoV, autonomous vehicles. Various built-in sensors in such devices collect the data for swift processing and analysis. Therefore, while moving toward 6G, we will be witnessing sensor data sharing (exchange of information about the environment) among autonomous vehicles and edge/cloud servers.

 The ability to interact by sharing data (information) among various nodes and continuously sense the dynamically varying conditions of the environment is one of the core drivers for vehicular (autonomous) networks. To realize the autonomous systems, different functionalities/subsystems need to converge. In the 6G era, communication and sensing will be tightly fused together to support multiple autonomous systems, e.g., UAVs, autonomous vehicles, and Industry 4.0 \cite{chowdhury20206g}. To this end, Wild et al. \cite{wild2021joint} focused on join communications and sensing (JCAS) design aspects for beyond 5G and 6G networks. The authors offered  key drivers for integrating communications and sensing (e.g., AI and signal processing, massive bandwidth, denser networks, and MIMO and beamforming), analyzed the waveform appropriate for communications and radar sensing, considered various techniques for converging communications and sensing capabilities, discussed visions for advanced communications and sensing systems built upon distributed MIMO, and presented many research challenges for JCAS, which should be overcome to attain natively integrated communications and sensing in the 6G mobile networks.

 Beyond ISAC or JCAS, 6G is expected to be the 1st generation of cellular networks where localization, sensing, communication, and computing functionalities will be tightly integrated. One of the global 6G initiatives is the Hexa-X flagship project that consolidates 25 major participants from academia and industry; among the explicit objectives of the Hexa-X project is researching fundamentally new RATs, high-resolution locations, and sensing. The focus of 6G is to merge physical, digital, and human worlds, and the bridge that connects these worlds is the capability to sense, localize, and track physical objects. Under Hexa-X lie the integral parts of 6G comprising vision, radar, localization, and sensing. This will offer the ultra-high performance required for supporting location accuracies and latencies foreseen in the recognized/identified use-case families; it will also result in the tight integration of communications, radar, computing, localization, and sensing at both physical and software levels \cite{wymeersch2021integration}.

 Recently, the integration of communication, computing, caching, control, and sensing gains considerable interest from academia and industry due to its strong potential in 6G networks. Chowdhury et al. \cite{chowdhury20206g} described that 5G largely overlooks the integration of sensing, communication, computing, intelligence, and control functionalities. 6G promises to fulfill this lagging and cope with the 5G constraints for accommodating new challenges. In particular, the requirements of emerging IoIT/IoE applications, including XR, haptics, telemedicine, automation, and robotics, will surpass the capabilities of 5G and necessitate the integration of communication, computing, caching, control, and sensing in the 6G and beyond networks \cite{chowdhury20206g,saad2019vision}. In other words, converging AR and VR cannot suffice the challenging requirements of several applications, such as near-real-person video conferencing, remote surgery/diagnosis, and ultra-high-resolution remote sensing for remote exploration. Today, holographic teleportation is considered as a potential replacement for AR/VR-enabled solutions\cite{akyildiz20206g}.  
 The holographic and high-precision communication for haptics and Tactile applications will be supported by the two fundamental drivers for 6G, viz., IoE and mobile Internet, to realize comprehensive sensory responses, i.e., hearing, vision, smell, touch, and taste. This calls for processing enormous data in near-real time, ultra-high throughput (about 1-5 Tbps), and ultra-low latency (about 1 ms). Hence, there is need for integrating sensing, control, communication, caching, and computing capabilities in the 6G networks. Based on intelligent control, the integration of these capabilities will enable the networks to optimally decide what objects need to be sensed, what computational tasks need to be processed by which computing resources, and what data need to be stored by which caching resources.
 
 However, realizing the integration of communication, computing, caching, control, and sensing encounters some difficulties due to: i) complex communication resources (e.g., multi-dimensional radio and x-haul resources), ii) multi-layered computing resources (e.g., x-computing), iii) multi-layered caching resources, and iv) a large quantity of sensing objects (e.g., environments, humans, and things) for verticals and IoE applications. Fortunately, AI emerges capable of choosing appropriate sensing objects and competently managing communication, computing, and caching resources through learning from data, training, predicting, and decision making \cite{zhang20196g}. Thus, incorporating AI into wireless network domains will efficiently overcome these challenges to achieve the integration of communication, computing, caching, control, and sensing.

\section{OPEN CHALLENGES AND FUTURE DIRECTIONS}

In the previous section, we present a plethora of research works pertaining to the i4C. In this section, we will introduce: A) a number of open challenges and B) future directions for potential efforts on the i4C. 

\subsection{Open Challenges}

Of course, the i4C can bring promising opportunities for future networks. Despite this bright future, the i4C raises several  challenges that call for proper handling prior to fully implement it in 6G and beyond networks. To this end, this subsection highlights some critical challenges pertaining to the i4C. 

\subsubsection{The 4C Tradeoff}	

In \cite{liu2016three}, it was proved that with an integrated system harnessing synergistic combinations of different functionalities/resources, the same types of services can be realized. Nevertheless, the tradeoffs amongst the intrinsic 4C functionalities/resources for each type of service have to be fully investigated separately against the associated performance metrics and the constraints of the functionalities/resources. Hence, the optimal tradeoff of the functionalities/resources for each separate case, which is obviously important to determine, remains a key challenge in this context; see \cite{wang2017integration}.

\subsubsection{Mobility in 4C}	

One of the key factors that may account for frequent channel disconnection between the mobile UEs and the network edges is mobility. Due to the dynamic characteristics of wireless network parameters, including jitter, bandwidth, latency, and so on, the QoS of an application can be deteriorated when a mobile UE happens to be in moving state \cite{Ahmed2016ASO}. 

Wang et al. \cite{wang2017survey} pointed out that user mobility seriously disrupts both caching and computation offloading decisions and results in recurring handovers among the servers of the network edges. In other words, the user mobility and the short coverage of network edges contribute immensely to: i) degrading the efficiency of wireless networks, ii) drastic reduction in users' QoS, and iii) interrupting ongoing edge services \cite{wang2018survey}. In short, frequent user mobility can severely limit the performance of the i4C, since joint consideration of communication channel and computation capacity is essential for offloading the computational tasks. To that end, mobility management becomes necessary and needs to be redesigned in 4C scenarios \cite{wang2018learning}.

\subsubsection{Interference in 4C}

Signal interference plays a critical role in the wireless communication channel. To be more specific, if there is interference between different UEs, the control signal may be lost, thereby giving rise to further problems, such as high energy consumption, transmission delay, and lower bandwidth. This implies an adverse consequences on the communication resource, on which the other three key resources may depend. Hence, the wireless channel interference remains a key challenge that requires further research investigations.

\subsubsection{Security Issues in 4C}

Today, 4C functionalities are mostly hosted at the close proximity of UEs (e.g., the network edges) to save both energy and time. One of the finest examples of such hosts are the MEC servers, and deploying such servers to the network edges amounts to exposing them to vulnerable security threats. Moreover, disrupting the severs by any physical or cyber threat is tantamount to disrupting the 4C services/functionalities, limiting the performance of the integrated system of 4C. Therefore, securing the 4C functionalities/services at the network edges becomes a technical challenge to be addressed.

\subsubsection{Data Privacy}

Data privacy is another critical challenge that quests for further investigations in the 4C scenarios. In the integrated system of 4C, substantial amount of data may be offloaded from UEs to the servers of the network edges for real-time analytics, processing, and caching services. Such data could be: i) metadata, such as geographical locations, timestamps, and so on; ii) computing strategies, which includes computation offloading strategy; and iii) monitoring data \cite{hassan2019edge}.

Indeed, the emergence of edge computing contributes to dramatic reduction of end-to-end latency; however, data privacy will more than likely face severe challenges. The threat of data tampering and information leakage is aggravated as a result of wireless channel properties for computation offloading. For instance, if an edge server happens to be under the control of a malicious eavesdropper, the enterprise multimedia data may be transmitted to the server under the control. In this respect, the multimedia data can be handily eavesdropped or even tempered by the eavesdropper. Presently, data encryption is employed to guarantee data privacy and security. Nevertheless, due to the wireless channel characteristics and high complexity in computation, which result in latency and low QoE, data encryption may not sufficiently address the issues of data privacy in 4C scenarios \cite{nie2020data}.

\subsubsection{Ultra-low Latency Requirements in 4C: }

Convergence of 4C has broad range of mission-critical applications, among which are the UAV flight control systems, Tactile Internet, XR applications, IoIT, IoE, autonomous vehicles, and Internet of vehicles. Such applications usually require extremely low latency in a few tens of milliseconds. Unfortunately, the conventional wireless systems cannot meet such low latency requirements, which in return create additional intense challenges for the integrated system of 4C; see \cite{wang2017integration}.

\subsection{Future Directions}

\subsubsection{5G, 6G, and Beyond Networks}

Nowadays, myriad intelligent applications and use cases, such as XR, IoE, IoIT, autonomous vehicles, blockchain, Tactile Internet, Telesurgery, et cetera, surface with different service requirements. 5G, having 4C functionalities interacting at the proximity of such applications, promises to handle their diverse QoS requirements. However, user devices and their intelligent applications keep proliferating exponentially with stringent requirements, driving the 5G capabilities to their limits. This necessitates researchers in academia and industry to look beyond the boundaries of 5G networks. To fully support the emerging intelligent applications and use cases in the coming decades, there is need to look forward to converging the 4C functionalities in 6G and beyond networks.

\subsubsection{Integration }

The recent emergence of edge intelligence will undoubtedly trigger further research investigation. The point is that various network functionalities/resources, including communication, computing, caching, and control, are involved in the architecture of the edge networks. Yet, a systematic convergence of 4C (with the capability of realizing the system-level (optimal) performance) is far from being concluded \cite{wang2017survey}. Besides, there is still an urgent need for more holistic and intelligent control schemes to optimally control, coordinate, and integrate the network edge functionalities. Thus, more mechanisms for the i4C functionalities are required at the proximity of UEs. On the other hand, several emerging applications and use cases necessitate the convergence of communication, computing, control, and sensing in the 6G era.  Suffice it to say, research on integration has to continue in the future.

\subsubsection{Fundamental Relations behind 4C}

Most of the existing research efforts conducted on the i4C focused on improving capacity, latency minimization, and energy savings in  mobile networks. Of course, there are many more promising solutions behind a fully integrated system of 4C. However, such system can be successfully realized by leveraging the full synergy, capabilities, and tradeoff of the 4C functionalities. Jiao et al. \cite{jiao2018proactive} argued that the fundamental benefits behind utilizing the caching and computing functionalities/resources in mobile networks have not yet been studied effectively. Another effort in \cite{andreev2016exploring} pointed out that the ultimate synergy behind a fully integrated solution of 4C is not nearly well understood. To this end, synergistic collaboration, tradeoffs, and capabilities of 4C need to be further studied in order to fully leverage and benefit from the integrated  solution of 4C in 6G and beyond networks.

\subsubsection{The Capacity of 4C}

Although communication capacity can be determined in terms of Shannon information transmission theorem, the theoretical capacity of each measure (dimension) of caching, computing, and control functionalities is not yet determined. Hence, determining the theoretical capacity of each individual functionality of 4C remains a potential direction that quest for further investigations. The classical information theoretical model (by which instantaneous rate regions is addressed) is not directly applicable to the converged system of these functionalities. This is because it lacks ability to efficiently address caching-induced non-causality in the system \cite{wang2017integration}, \cite{liu2016three}. To sum up, an effort in \cite{liu2014content} uncovered the insufficiency of the conventional concept of rate capacity to portray the network strength (ability) to deliver (release) non-private contents for several UEs. The effort, which aimed at measuring the impact of caching in content releasing, introduced a so-called content rate for measuring the rate at which the amount of cached data is released to UEs through a shared wireless channel.

The explicit role that computing plays in the integrated system capacity measurement and calculation is not properly understood, despite numerous existing investigations devoted to integration of wireless communications and computing. A typical approach that has to do with this is network coding, which distinguishes algebraic operation (computing) from communication operation. Other approaches, such as collaborative transmission \cite{gesbert2010multi} and distributed MIMO \cite{you2010cooperative}, do not decouple computing and communication, as logical operations and channel coding across information streams are entwined. Obviously, a bound together capacity analysis that portrays communication, computing, and caching resources in canonical frame, bears considerable hypothetical value, and as such, is in critical interest; see \cite{wang2017integration,liu2016three}.

\subsubsection{Real-Time Decision Making}

Due to their proximity to UEs, MEC and other edge networks can track their real-time information like user's location, behavior, and resources environment \cite{wang2017survey},\cite{mao2017survey}. Delivering context-aware services to UEs can be enabled by inference based on such information. For example, for video guidance in a museum, the subscribers' interests can be predicted/learnt through an AR application, based on their (subscribers') positions in the museum, for delivering contents, such as artworks and antiques. The CTrack system is another example, which tracks and predicts multiple subscribers' trajectories by using BS fingerprints for routing, navigation, monitoring, and personalized trip management \cite{mao2017survey}. The success behind realizing such real-time control systems requires deep insights into the communications, computing control \cite{xia2004integrated} and caching theories. On top of that, proactive resource allocation requires the use of different levels of real-time information, i.e., application, network, and UE levels \cite{wang2017survey}.

\subsubsection{Intelligence for the  i4C}

Recent research investigations resort to intelligence to overcome critical issues of i4C. This may not be unconnected with the complexity of the integrated system and specific use cases in MEC. The study in \cite{wang2020convergence} raised some key points behind considering intelligence for resource optimization/integration. As discussed in the sequel, AI frontiers, such as ML, DRL and FL, will have pivotal roles to play in 4C scenarios.

\textit{\textbf{Machine Learning:}} The 5G/6G networks will be endowed with the swarm intelligence and node intelligence to improve their efficiency. Against conventional objective function of a single component, a trade-off bounded by various factors, such as energy consumption, delay, capacity, complexity, and so on, will be dealt with for management and allocation of resources \cite{wang2020thirty}. Hence, heterogeneous network devices, different QoS demands, as well as large state and action spaces will greatly complicate the  i4C in 5G, 6G, and beyond networks. In the face of such complication, future wireless networks may rely on ML for online and/or fully-distributed algorithms. Moreover, ML has capability of dealing with the challenges of closed-form solution, problem formulation, and other issues of channel modeling inflicted by model-free wireless networks \cite{pham2020survey}. In fact, by leveraging learning based on try and error experiments, ML is found efficiently capable of handling multi-objective optimization problems, especially in light of managing the multi-agent collaborative networks \cite{wang2020thirty}.

Needless to say, DRL will potentially play major roles in breaking the complexity of join optimization of 4C especially in extremely time-varying MEC systems. DRL emerges through the coupling of reinforcement learning algorithm with deep learning to combat considerable input data amount and determine the optimal policy for the complicated resource allocation problems. In DRL, deep Q-network (DQN) can be leveraged for approximating the Q value function \cite{wang2017integration}. In \cite{li2018deep,hu2019ready}, Q learning and DQN have been studied to overcome some challenges related to joint optimization of 4C; see \cite{pham2020survey}. Likewise, \cite{he2018trust,he2017integrated,huang2019deep} applied DRL approach to investigate the integration of communication, computing, and caching.

\textit{\textbf{Federated Learning:}} Data privacy is a critical challenge that calls for more attention. In contrast to conventional approaches of ML, which have no room for preserving the privacy of training data, FL can enable UEs to collaboratively learn a shared model while locally keeping their respective data. In other words, FL can allow the distribution of training data across individual UEs. Hence, the limitations of distributed learning can be dealt with. Such limitations include: i) low efficiency caused by heterogeneous capabilities in UEs and network states, ii) insufficient time and training data, iii) lopsidedness in the number of the training data samples, as well as iv) non-independent and identically distributed data between the UEs.

In a computation offloading technique, a significant number of UEs may offload their tasks for remote computation and caching. In tradition, the UEs decide whether to offload the tasks or not by reporting their individual information, which involves battery lifespan, channel gain, computing capabilities, and so on, to the network edge. Unfortunately, malicious eavesdroppers can access and even use the information illegally to obtain the locations of the UEs. Applying FL will allow each individual UE to download the master model from the network edge and thereby learn the computation offloading decisions based on its local information only. Moreover, based on the updates obtained from the UEs, the network edge will be responsible for the master model update. In this way, FL will preserve the privacy of data and bring distributed offloading decisions. In summary, FL can be considered as one of the finest potential solutions to resource allocation/integration issues. Thus, it is foreseen to serve as a sharp tool for various challenges related to resource allocation optimization in MEC \cite{pham2020survey}.

\section{CONCLUSION}

This article brings an extensive survey on the i4C, which becomes indispensable due to recent growth of smart devices and their emerging mission-critical applications. The survey starts with providing a snapshot of different aspects of the i4C, including motivations, some potential applications and use cases, and key enabling technologies. To lay the foundation of the integration in 5G, 6G, and beyond networks, the article offers a firsthand tutorial on various models of 4C. Then, it reviews several state-of-the-art efforts on the i4C, placing emphasis on recent trends of conventional optimization and AI-based integration approaches. It disscuses the convengence of communication and sensing and classifies different approaches of resource integration. Finally, open challenges and future directions are discussed.

\bibliographystyle{IEEEtran}
\bibliography{myref}
\newpage
\biographies
% \section*{Biographies}
%\begin{CCJNLbiography}{photo.eps}{Bei Liu}
%received the B.S. and Ph.D. degrees from Beijing University of Posts and Telecommunications (BUPT) in 2001 and 2006, respectively. He is currently %an Associate Professor in School of Electronic Engineering, BUPT. His %current research inter-ests include wireless communication theory and %technology, wireless mesh networks.
%\end{CCJNLbiography}

%\begin{CCJNLbiography}{photo.eps}{Fei Zhang}
%received the B.S. degree in electronic science and technology from %Beijing University of Posts and Telecommunications (BUPT) in 2017. He is %currently working towards the master degree in School of Electronic %Engineering, BUPT. His current research is signal processing technology %for wireless communications.
%\end{CCJNLbiography}

\end{document}